\documentclass[a4paper,fleqn]{cas-sc}

\definecolor{cite_color}{rgb}{0.0, 0.58, 0.71}
\usepackage{natbib}
\setcitestyle{authoryear,open={},close={}} 
\usepackage{hyperref}
\hypersetup{colorlinks=true,citecolor=cite_color,linkcolor=cite_color}
\usepackage{amsthm}
\usepackage{subcaption}
\usepackage{lineno}
\usepackage{xcolor}
\definecolor{db}{rgb}{0.0, 0.2, 0.7}

\renewcommand{\figurename}{Fig.}

\makeatletter
\renewcommand*{\fnum@figure}[1]{\figurename~\thefigure.}
\makeatother
\def\tsc#1{\csdef{#1}{\textsc{\lowercase{#1}}\xspace}}
\tsc{WGM}
\tsc{QE}
\tsc{EP}
\tsc{PMS}
\tsc{BEC}
\tsc{DE}

\begin{document}
\let\WriteBookmarks\relax
\def\floatpagepagefraction{1}
\let\printorcid\relax 

\def\textpagefraction{.001}
\shorttitle{}
\shortauthors{Mohammad Anis et~al.}

\title [mode = title]{Real-time risk estimation for active road safety: Leveraging Waymo AV sensor data with hierarchical Bayesian extreme value models}

\author[1]{\textcolor{black}{Mohammad Anis}}[]

\credit{Conceptualization, Methodology, Writing – original draft,  Software, Writing – review \& editing}

\address[1]{Zachry Department of Civil $\&$ Environmental Engineering, Texas A$\&$M University, College Station, TX 77843, USA}

\author[1]{\textcolor{black}{Sixu Li}}[]
\credit{Writing – original draft,  Writing – review \& editing}

\author[2]{\textcolor{black}{Srinivas R. Geedipally}}[]
\credit{Writing – review \& editing}

\address[2]{Center for Transportation Safety, Texas A$\&$M  Transportation Institute, 111 RELLIS Parkway
Bryan, TX 77807, USA
}

\author%
[1]
{\textcolor{black}{Yang Zhou}}
\cormark[1]
\ead{yangzhou295@tamu.edu}
\credit{Conceptualization, Methodology, Writing – review \& editing, Supervision}

\author%
[1]
{\textcolor{black}{Dominique Lord}}
\credit{Conceptualization, Methodology, Writing – review \& editing, Supervision}

\cortext[cor1]{Corresponding author}

\begin{abstract}
Near-miss traffic risk estimation using Extreme Value Theory (EVT) models within a real-time framework offers a promising alternative to traditional historical crash-based methods. However, current approaches often lack comprehensive analysis that integrates diverse roadway geometries, crash patterns, and two-dimensional (2D) vehicle dynamics, limiting both their accuracy and generalizability. This study addresses these gaps by employing a high-fidelity, 2D time-to-collision (TTC) near-miss indicator derived from autonomous vehicle (AV) sensor data. The proposed framework uses univariate Generalized Extreme Value (UGEV) distribution models applied to a subset of the Waymo motion dataset across six arterial networks in San Francisco, Phoenix, and Los Angeles. Extreme events are identified through the Block Maxima (BM) sampling-based approach from each conflicting pair, with 20s block sizes to account for the scarcity of samples in short-duration traffic segments. The framework also incorporates conflicting vehicle dynamics (e.g., speed, acceleration, and deceleration) as covariates within a non-stationary hierarchical Bayesian structure with random parameters (HBSRP) UGEV models, allowing for the effective management of vehicle spatial, temporal, and behavioral heterogeneity. Results show that HBSRP-UGEV models outperform other approaches, with a 6.43-10.56\% decrease in DIC, especially for near-miss events in short-duration traffic segments. The inclusion of dynamic vehicle behaviors and random effects substantially enhances the model's capability to estimate real-time traffic risks. This generalized real-time EVT model bridges the gap between active and passive safety measures, offering a precise and adaptable tool for network-level traffic safety analysis.

\end{abstract}


\maketitle

\section{Introduction}

Ensuring highway safety is critical for achieving Vision Zero and upholding the integrity of our transportation system. In 2021, there were 39,508 fatal crashes in the United States, with a fatality rate of 1.37 per 100 million vehicle miles traveled (VMT) (\citep{NHTSA}). Additionally, these crashes impose an economic burden of approximately \$498.3 billion annually (\citep{nsc}). Such alarming statistics underscore the need for practical solutions to enhance road safety. 

The emergence of autonomous vehicles (AVs) has the potential to significantly enhance highway safety by improving perception, reacting promptly to ambient traffic, and reducing driver distraction and fatigue (\citep{dai2023explicitly}). However, in current conditions, human-driven vehicles (HDVs) will coexist with AVs for the foreseeable future; understanding driving behavior remains essential to improving road safety.

Previous studies have identified human driving behavior as a significant contributor to vehicle crashes (\citep{zhu2023investigation,yue2018assessment,antin2011design,klauer2006impact,petridou2000human,rowe2015measuring}), approximately 94\% attributed to driving errors  (\citep{singh2018critical}). Critical driving behaviors, including speed, acceleration, deceleration, braking, steering angle, etc. (\citep{li2019drivers}), directly impact vehicle dynamics and are essential for assessing crash risk. Analyzing individual vehicle dynamics is critical for reducing crash frequency; this approach has yet to be fully explored. Past research primarily focused on average vehicle dynamics over an observation period. This may obscure critical analysis of heterogeneity in driving behavior over a short duration, particularly in real-time crash estimation frameworks. While driving simulators provide some insight, they often fail to capture the complexity of real-world driving. As transportation networks grow more complex, ensuring driver safety demands thorough safety analysis and investment to save lives and reduce economic losses.

Traditional safety analysis often relies on historical crash data to predict future extreme events or crashes within specific roadway entities (intersections, midblocks, corridors, etc.), widely known as the reactive approach (\citep{arun2021systematic,lee2017intersection,pei2011joint}). However, this approach has significant drawbacks (\citep{lord2021highway}), including the rarity and randomness of crashes, the lengthy data collection process  (\citep{mannering2020big}), limited sample sizes, unobserved heterogeneity (\citep{lord2010statistical}), and inherent issues: under-reporting and subjective reporting (\citep{arun2021systematic, tarko2018estimating}). These limitations hinder the accuracy of crash risk estimates, reliable safety inferences, and comprehensive safety assessments (\citep{tarko2018surrogate,mannering2016unobserved}), thereby stalling advancements in highway safety research (\citep{wang2021review}). To overcome these limitations, the near-miss analysis offers a proactive alternative (\citep{ali2023assessing,arun2021systematic,mahmud2017application}). Near-miss events are more frequent than crashes, providing an enriched dataset for robust analysis and valuable insights into vehicle interactions and road network performance (\citep{tarko2018estimating,davis2011outline}). Surrogate safety measures (SSMs) are commonly used to identify near-misses from road users' trajectories, enabling more proactive safety assessments. Usually, the trajectory extraction relied on infrastructure-based technologies (e.g., roadside cameras, loop detectors) and movable systems (e.g., lidars, radars, and computer visions). These systems can monitor and report traffic states (e.g., speed, acceleration, brake, vehicle dimensions, etc.) for all individual vehicles within their detection range (\citep{islam2021crash, yuan2019real,li2020real}). Infrastructure-based systems face occlusion, shadowing, overlap detection, and maintenance challenges, which affect data accuracy (\citep{st2013automated}). In contrast, autonomous vehicle (AV) sensors provide high-resolution, granular, real-time data on individual vehicle dynamics, effectively overcoming these limitations. However, this granular data is often underutilized in real-time near-miss risk estimation frameworks. Some studies in this domain have primarily focused on specific near-miss types (rear end, lane changing, overtaking, angle collision) or combining analyzing at certain locations or networks (\citep{kamel2023real, kamel2024real,ghoul2023dynamic,ghoul2023real}). These studies lack a comprehensive framework accommodating diverse roadway geometry, locations, and various near-miss types, limiting the findings' generalizability.

Precise near-miss event extraction (considered precursors of crashes) is critical for estimating the precise crash risk because of their inherent relationship. Recently, this relationship (near miss and real crashes) has garnered significant attention in previous research (\citep{ali2023assessing,zheng2021validating,songchitruksa2006extreme}). Several methods explored to infer this relationship accurately, including extreme value theory (EVT) (\citep{zheng2014freeway, tarko2012use,songchitruksa2006extreme}), probabilistic frameworks (\citep{saunier2008probabilistic}), and causal models (\citep{davis2011outline}). EVT has been widely adopted in recent years since its excellent ability to extrapolate from frequently occurring (near-misses) events to unobserved (crashes) events (\citep{coles2001introduction}). Songchitruksa and Tarko (\citeyear{songchitruksa2006extreme}) pioneered EVT-based crash estimation, and recent extensions by Zheng et al. (\citeyear{zheng2019univariate,zheng2020novel}) introduced Bayesian hierarchical structures that account for non-stationarity and unobserved heterogeneity, making EVT ideal for real-time safety analysis. 

Real-time safety analysis leverages near-miss events collected over short intervals to estimate crash risk, thus supporting proactive traffic safety management  (\citep{zheng2020novel,fu2022bayesian}). However, previous studies often rely on sparse data, which limits their ability to capture temporal variations and restricts their applicability for network-wide safety assessments. Recent studies have applied Bayesian hierarchical spatial random parameter EVT models to handle sparse data (\citep{kamel2023real, kamel2024real}). However, these studies have focused on one-dimensional (1D) or longitudinal near-miss events along roadways without considering two-dimensional (2D) vehicle dynamics.

This study proposes a generalized real-time traffic risk estimation framework that uses a high-fidelity, 2D near-miss indicator (beyond specific crash patterns), extreme events are integrated into hierarchical Bayesian structure univariate Generalized Extreme Value (UGEV) distribution models to estimate traffic risks across diverse highway geometries. Near-miss events are derived from high-resolution Waymo motion data, updated every 0.1s, across six locations in three major cities: San Francisco, Phoenix, and Los Angeles. Extreme events are identified through the Block Maxima (BM) sampling-based approach from each conflicting pair; each block size is 20s, which effectively addresses the scarcity of samples in short-duration traffic segments. The framework also incorporates conflicting vehicle dynamics as covariates into non-stationary hierarchical Bayesian structure with random parameters (HBSRP) UGEV models, effectively managing unstable distributions in different sites, limited extreme events, and spatial or
temporal driving behavior heterogeneity. This framework overcomes the limitations of prior studies and offers more accurate traffic risk assessments. Results demonstrate that this framework enhances understanding of the relationship between vehicle dynamics and traffic risk, improving the accuracy of proactive safety assessments. This study aims to fill existing research gaps by focusing on the following objectives:

\begin{itemize}

\item Develop a generalized real-time traffic risk estimation framework using a high-fidelity, 2D near-miss
indicator that accommodate diverse roadway geometries and crash patterns.

\item Introduce a conflicting pair approach for Block Maxima (BM) based sampling (each block size is 20s), effectively addressing the scarcity of samples in short-duration traffic segments.

\item Incorporates conflicting vehicle dynamics (e.g., speed, acceleration, and deceleration)
as covariates into non-stationary hierarchical Bayesian structure with random parameters (HBSRP) UGEV models, allowing for the effective management of vehicle spatial, temporal, and behavioral heterogeneity.

\end{itemize}

The structure of this study is as follows: In Section \ref{sec2}, we present a comprehensive review of the literature relevant to the proposed framework. Section \ref{sec3} covers the data description, a generalized real-time risk estimation EVT framework based on 2D high-fidelity near-miss risk indicator TTC. In Section \ref{sec4}, we discuss data preparation, model estimation inference, and model validation in detail. Finally, a brief discussion is included in Section \ref{sec 5}, and we draw a summary and conclusions from our study in Section \ref{sec 6}.

\section{Literature review}\label{sec2}
Transportation safety research has traditionally relied on crash data to guide interventions and mitigate future risks. However, the reactive nature of these methods limits their ability to effectively reduce crash risk, as discussed by several studies (\citep{lord2021highway, mannering2020big, arun2021systematic}). To overcome these challenges, proactive safety approaches such as near-miss analysis have gained traction. Unlike crashes, near-misses are much more frequent, providing a richer dataset for understanding vehicle interactions. Abdel-Aty et al. (\citeyear{abdel2010concept, abdel2023real}) discuss using real-time data to identify potential crash locations and deploy traffic management strategies. Other studies (\citep{yang2021proactive,yang2022functional}) propose using traffic cameras to analyze driver response behavior, detect anomalies, and provide real-time warnings. Tarko (\citeyear{tarko2018surrogate}) and Davis et al. (\citeyear{davis2011outline}) advocate for using SSMs to quantify and derive these near-miss events, enabling active safety interventions.

SSMs are commonly used to quantify near-miss events. These measures help derive meaningful safety metrics by evaluating temporal and spatial proximities between road users, which aids in detecting, evaluating, and assessing the severity of near-miss events (\citep{abdel2023advances}). SSMs have facilitated infrastructure assessment, user behavior analysis, and the evaluation of new technologies (\citep{oikonomou2023conflicts}). Within the SSM framework, three primary sub-categories exist: time-based, deceleration-based, and energy-based measures. Time-based SSMs play a crucial role in highway safety research, with TTC being one of the most widely used examples, first utilized by Hayward (\citeyear{hayward1971near}). TTC estimates the time remaining before a collision by analyzing projected vehicle paths, assuming steady speed and direction, and using probabilistic models to gauge a driver’s likelihood of avoiding a crash (\citep{wang2014evaluation}). However, due to the limitations of TTC in scenarios where drivers frequently perform evasive maneuvers, Perkins and Harris (\citeyear{perkins1967traffic}) developed Time to Accident (TA). Extending TTC, Minderhoud and Bovy (\citeyear{minderhoud2001extended}) introduced more sophisticated SSMs, such as Time-Exposed Time to Collision (TET) and Time-Integrated Time to Collision (TIT). TET quantifies the duration a vehicle spends below a critical TTC threshold, while TIT measures the area between the TTC curve and this threshold under hazardous conditions, requiring continuous TTC monitoring. Additional time-based SSMs include time headway (\citep{vogel2003comparison}), time to zebra (TTZ) (\citep{varhelyi1998drivers}), time-to-lane crossing (TTL), modified time to collision (MTTC) (\citep{ozbay2008derivation}), and post-encroachment time (PET) (\citep{allen1978analysis}). MTTC adjusts for variable speeds in car-following situations. In contrast, PET assesses the interval between one vehicle leaving and another entering a potential conflict point, independent of assumptions about speed or direction. Similarly, Venthuruthiyil and Chunchu (\citeyear{venthuruthiyil2022anticipated}) proposed a novel surrogate safety indicator called anticipated collision time (ACT) that captures various crash risk patterns proactively. Despite their extensive application, time-based SSMs have limitations, such as difficulty capturing interactions involving angled approaches and lateral maneuvers. They are also challenged in representing complex, real-world traffic dynamics in multi-directional interactions

The emergence of real-time traffic safety analysis has shifted the focus toward leveraging continuous data to assess and mitigate risks as they evolve (\citep{yuan2018utilizing}). Studies have shown that real-time traffic factors such as average speed, acceleration, upstream volume, etc., significantly impact crash occurrence (\citep{yuan2018real}). Deep learning techniques and UAV-based video analysis have also been proposed for real-time traffic analysis, providing a potential solution for efficiently managing urban traffic (\citep{zhang2019real}). The relationship between crash occurrence and real-time traffic characteristics can be analyzed through Bayesian conditional logistic models (\citep{yuan2018real}). In recent years, EVT has been increasingly applied to traffic safety analyses, particularly in estimating rare and severe events, offering quick and reliable evaluations without relying on historical crash data (\citep{orsini2019collision}).  Orsini et al. (\citeyear{orsini2020large}) and Zheng et al. (\citeyear{zhang2019real}) both demonstrate the effectiveness of EVT in predicting road crashes using risk indicators, explicitly focusing on rear-end collisions and exploring the use of univariate and bivariate EVT models. Fu et al.(\citeyear{fu2021random,fu2022random}) further enhance the application of EVT by proposing a random parameters Bayesian hierarchical modeling approach to account for unobserved heterogeneity in near-miss extremes. This approach outperforms traditional models in terms of crash estimation accuracy and precision. EVT has also been used in before-after road safety analysis, which has shown the capability to confidently estimate extreme events and identify safety improvements (\citep{zheng2019application}). Recent studies have highlighted the importance of incorporating vehicular heterogeneity in crash risk assessment (\citep{kumar2024risk}). A dynamic approach to identifying hazardous locations using a conflict-based real-time EVT model has been proposed, allowing for assessing short-term and longer-term crash risk (\citep{ghoul2023dynamic}). For real-time near-miss safety analysis, a Bayesian dynamic extreme value modeling approach has been developed that considers changes in time and non-stationary extremes (\citep{fu2022bayesian}). Zheng (\citeyear{zheng2014freeway}) and Wang et al. (\citeyear{wang2019crash}) both found that EVT models outperformed traditional statistical models in predicting crash frequency and probability. However, Ali (\citeyear{ali2023assessing}) highlighted the need for ongoing evaluation and development of EVT models, particularly in the context of AV. Research on the Block Maxima (BM) and Peak Over Threshold (POT) methods in EVT has yielded mixed results. The BM method, known for its simplicity and effectiveness in handling time-series data, is often preferred in real-time traffic conflict studies (\citep{fu2022bayesian}). This method is particularly advantageous due to its ability to handle temporal dependencies, traffic flow variations, and computational efficiency. The BM method has also developed a Bayesian dynamic extreme value modeling approach for real-time safety analysis based on conflict. This approach considers changing model parameters and conflict extremes that remain unchanged over time (\citep{fu2022bayesian}). Non-stationary BM based sampling are beneficial in capturing trends and variations in traffic conditions (\citep{mannering2020big}). Incorporating vehicle-specific characteristics as covariates in EVT models can enhance the accuracy of safety evaluations by addressing unobserved heterogeneity (\citep{kumar2024risk}). Kumar (\citeyear{kumar2024risk}) and Fu (\citeyear{fu2021random}) both emphasize the importance of considering vehicle dynamics such as speed, acceleration, and braking patterns in these models. Fu (\citeyear{fu2021random,fu2022random}) further suggests that using random parameters in Bayesian hierarchical EVT models can improve crash estimation accuracy and precision. These models can include random parameters and hierarchical structures, which capture the complexity of traffic systems and consider changes in time and non-stationarity. Kamal et al. (\citeyear{kamel2023real, kamel2024real}) extend this approach to real-time safety analysis using AV sensor data, demonstrating the ability of Bayesian hierarchical models to address the scarcity and non-stationarity of conflict extremes. 

\section{Methodology} \label{sec3}

\subsection{Data description \label{sec3.1}}

The Waymo Open Dataset, collected by a Society of Automotive Engineer (SAE) Level 4 AV in 2019, is a rich resource for analyzing AV and other road user movements. This dataset, generously provided by Waymo, has been anonymized to protect road user privacy by removing specific geolocation, date, and time information. It has been widely utilized in recent research for its high-resolution sensor data, which includes 3D point clouds, high-definition images, and precise localization information. The dataset contains over 100,000 scenes collected across six cities in the United States, capturing diverse scenarios, including varying weather conditions, times of day, and a mix of urban, suburban, and rural settings (\citep{ettinger2021large}).  Waymo AVs have collectively traversed over 32 million kilometers, and their sensors—LiDAR, cameras, and radar—provide data at 10 Hz, capturing detailed vehicle movements and the surrounding environment. The dataset includes sensor data such as 3D point clouds, high-definition images, and precise localization information, offering a comprehensive view of real-world driving conditions.

\begin{figure}[h]
    \centering
    \setlength{\abovecaptionskip}{0pt}  \includegraphics[width=0.7\textwidth]{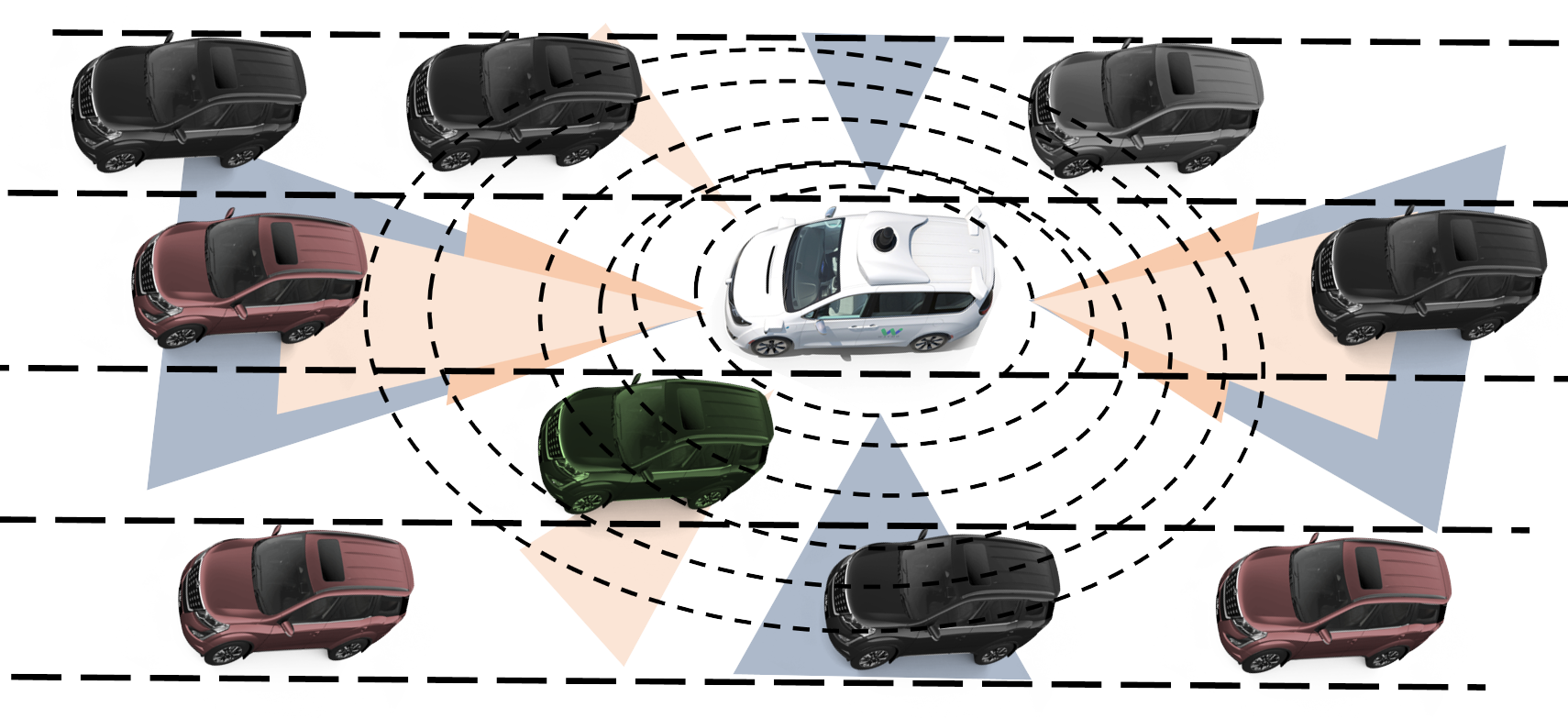}
    \caption{Waymo AV and its ambient environment}
    \label{fig:1}
\end{figure}

 Waymo's AVs are easily recognizable by their distinct technological features, which include prominently protruding cameras, a robust exterior frame, Waymo-branded stickers, and roof-mounted LiDAR sensors (Fig. \ref{fig:1}). These features are integral to the AVs' ability to perceive and navigate complex environments. The dataset also encompasses diverse scenarios, including varying weather conditions, different times of day, and a wide range of urban, suburban, and rural settings. Notably, 96.5\% of the data points were generated in urban environments, which also aligns with the focus of this study on urban roadways. Hu et al. (\citeyear{hu2022processing}) have highlighted the superior quality of this dataset compared to NGSIM.  These qualities in data make the Waymo open dataset an invaluable resource for advancing research in driving behavior, environmental interactions, and the development of robust traffic risk models for improving road environment performance. The dataset employed in this study is derived from the Mendeley open-source repository (\citeyear{Waymo_open_data}). This dataset has been meticulously processed and enhanced, with particular emphasis on paired car-following trajectories, thereby augmenting its utility for driving behavior research.

The dataset comprises two main subsets: perception and motion dataset. This study focuses exclusively on the motion dataset. This subset includes various dynamic and static features, such as the GPS location (modified), heading direction, speed, acceleration, steering angle, and volume, among other relevant information. A detailed summary of the raw data features is listed in Table \ref{Table:1}. The data are captured with a high temporal resolution, having a sampling rate of 0.1 seconds. This fine-grained sampling allows for precise tracking of vehicle dynamics. Each data segment covers 20 seconds, resulting in a total of 200 frames per segment. This high-frequency sampling and comprehensive feature set enable detailed vehicle motion and behavior analysis over short intervals. The dataset's distinguishable appearance and extensive feature set allow surrounding HDV to recognize and interact with HDV effectively. This interaction is crucial for studying the interaction patterns and safety aspects of AVs and HDVs in mixed-traffic environments.

\begin{table}[h]
\centering
\caption{Features in available Waymo AV open dataset}
\label{Table:1}
\begin{tabular}{l l l }
\hline
Name & Description & Unit \\ \hline
id & Unique ID for each data point & \\ 
segment id & Unique segment ID for each data point & \\ 
frame label & Unique Frame ID for each data point & \\ 
time of day & Day or night & \\ 
location & Three specific locations: Px, SF, and LA & \\
weather & Sunny, cloudy or dark & \\ 
laser veh count & Total vehicle count number of each frame & \\
obj type & Vehicle, pedestrian or cyclist & \\
obj id & Unique ID for each object & \\
global time stamp & Global time stamp & \\
global center x & Object global North-South position & \\
global center y & Object global East-West position & \\
global center z & Object global z coordinates & \\
length & Object length & $m$ \\
width & Object width & $m$ \\
height & Object height & $m$ \\
heading & Direction of the object at the instant the data point was captured & $rad$ \\
speed x & Speed of the vehicle along x axis at the instant the data point was captured & $m/s$ \\
speed y & Speed of the vehicle along y axis at the instant the data point was captured & $m/s$ \\
acce x & Acceleration of the vehicle along x-axis at the instant the data point was captured & $m/s^2$ \\
accel y & Acceleration of the vehicle along y axis at the instant the data point was captured & $m/s^2$ \\
steering angle & vehicle steering angle at the instant the data point was captured & $rad$ \\
\hline
\end{tabular}
\end{table}

\subsection{2D high-fidelity near-miss risk indicator \label{sec3.2}}

\begin{figure}[h]
    \centering
    \setlength{\abovecaptionskip}{0pt}
    \includegraphics[width=0.7\textwidth]{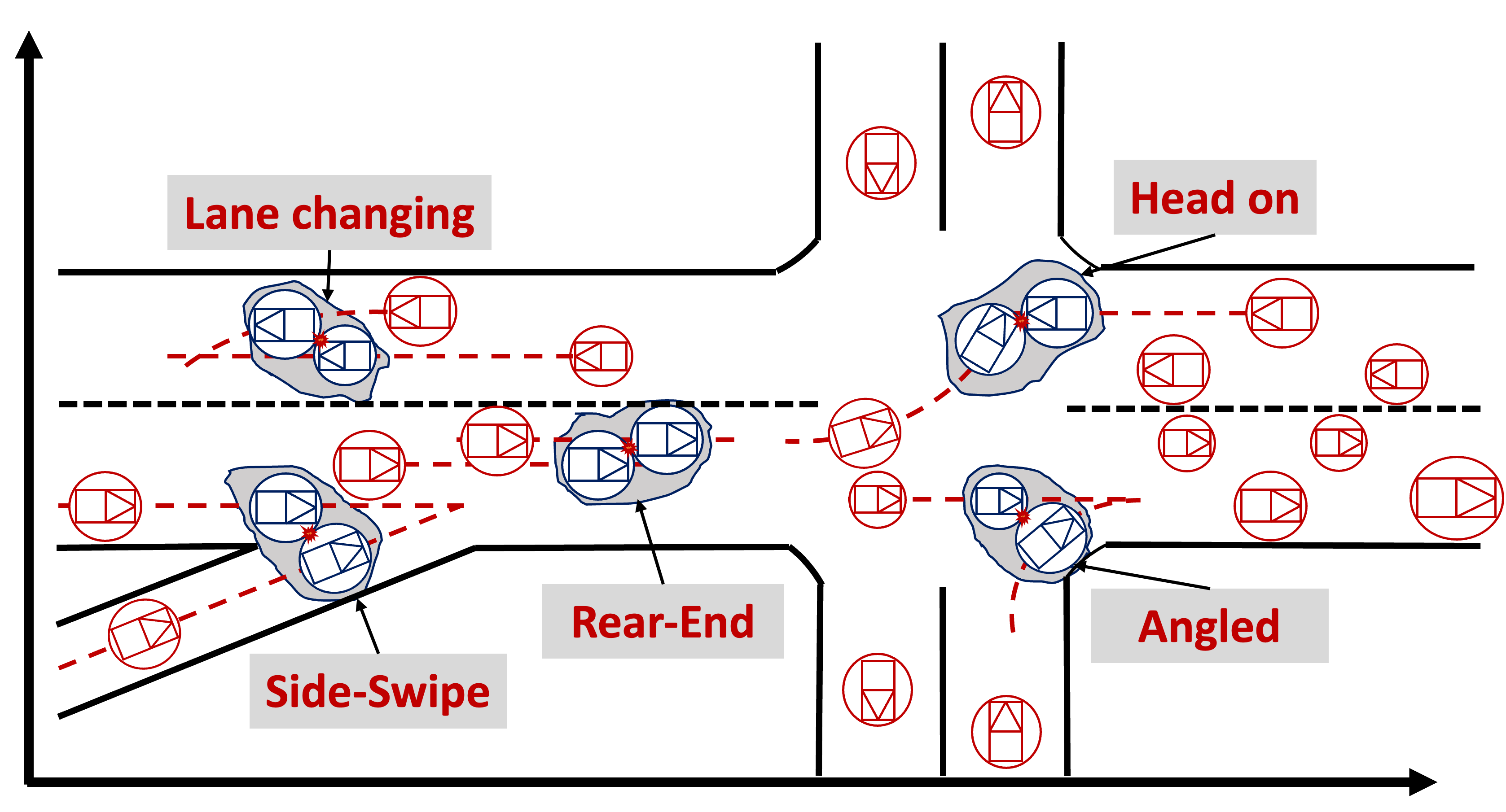}
    \caption{Representation of different types of collisions}
    \label{fig:2}
\end{figure}

The 2D high-fidelity near-miss risk indicator is introduced to address the limitations of traditional SSMs that primarily focus on 1D scenarios, capturing only longitudinal vehicle movements. In real-world traffic, however, vehicle interactions often involve complex maneuvers such as lane changes, merging, and evasive actions, requiring the consideration of both longitudinal and lateral dynamics. While some studies have addressed both dimensions, there are concerns about the accuracy and objectivity of using existing SSMs to characterize various conflicts (see Fig. \ref{fig:2}). It highlights the need for a versatile, high-fidelity analytical framework to accurately portray complex traffic potential collisions.

The 2D high-fidelity TTC derived by Li et al. (\citeyear{li2024beyond}), which determines potential collisions based on both longitudinal and steering movements of vehicles, is adopted.  Assumptions regarding acceleration and steering angle invariance are made each time the TTC is calculated. Under this assumption, the future trajectory of a vehicle is projected using the following state-space model in the equation: 

\begin{equation}
\dot{\chi_i}(t) = f(\chi_i(t), u_i(t)) \label{equation 1}
\end{equation}

Let \(i \in V\), where \(V\) is the set of vehicle indices. The vehicle state at time \(t\), \(\dot{\chi}_i(t) \in \mathbb{R}^n\), includes position and direction, while the control input \(u_i(t) \in \mathbb{R}^m\) covers acceleration and steering. Here, \(n\) and \(m\) represent the number of state variables and control inputs. Assuming \(t = 0\) as the start time, the future trajectory \(\chi_i(t)\) for \(t > 0\) can be derived from the initial state \(\chi_i(0)\) and control input \(u_i(t)\). Given the Eqn. (\ref{equation 1}), remains time-invariant and linearized future trajectories by simplifying the complex dynamics using a Taylor series expansion. This yields a linearized state-space representation, which is expressed in the form:

\begin{equation}
 \dot{\chi}(t) = A\chi(t) + Bu(t)+C \label{LTI vehicle model}
\end{equation}

Here, A, B, and C are matrices representing the system dynamics after linearization, capturing how the state evolves under the influence of control inputs.

To determine the TTC, the future state of the vehicles must be predicted. This is achieved by solving the linearized state-space system using the following matrix exponential solution:

\begin{equation}
\chi(t) = e^{At} \chi(0) + \int_0^t e^{A(t-\tau)}Bu(\tau)+e^{A(t-\tau)}Cd\tau \label{LTI solution}
\end{equation}

The term $e^{At}$ is computed using a matrix exponential, which can be represented as a Taylor series:

\begin{equation}
e^{At} = I_n + At + \frac{1}{2!}A^2 t^2 + \frac{1}{3!}A^3 t^3 + \dots + \label{exponential taylor series}
\end{equation}

This solution provides an analytical prediction of the future trajectory of each vehicle, incorporating both free responses (resulting from initial conditions) and forced responses (due to applied control inputs like acceleration and steering). For a deeper understanding, readers are directed to (\citep{li2024beyond}). To extend the proposed framework beyond 1D, Li et al. (\citeyear{li2024beyond}) employ a 2D kinematic bicycle model to describe vehicle movements in Cartesian coordinates. This model accounts for the vehicle's motion in both the x and y directions, providing a comprehensive representation of its dynamics for future state estimation using  Eqs. (\ref{eq1})-(\ref{eq4}). The model designate the state vector $\chi(t)=[x(t), y(t), \theta(t), v(t)]^T$ and the control input vector $u(t)=[\delta(t), a(t)]^T$. 

\begin{equation}
{x}=v\text{~cos}(\theta) \label{eq1}
\end{equation}

\begin{equation}
{y}= v\text{~sin}(\theta) \label{eq2}
\end{equation}

\begin{equation}
{\theta}=\frac{v\text{~tan}(\delta)}{L} \label{eq3}
\end{equation}

\begin{equation}
{v}=a \label{eq4}
\end{equation}

Here, $(x, y)$ represent the Cartesian coordinates of the vehicle's center of gravity (C.G.), $\theta$ is the vehicle heading angle, $\delta$ are the steering angles, $L$ is the vehicle's wheelbase, and $v$ and $a$ are the velocity and acceleration, respectively.

\begin{figure}[h]
    \centering
    \setlength{\abovecaptionskip}{0pt}
    \includegraphics[width=0.5\textwidth]{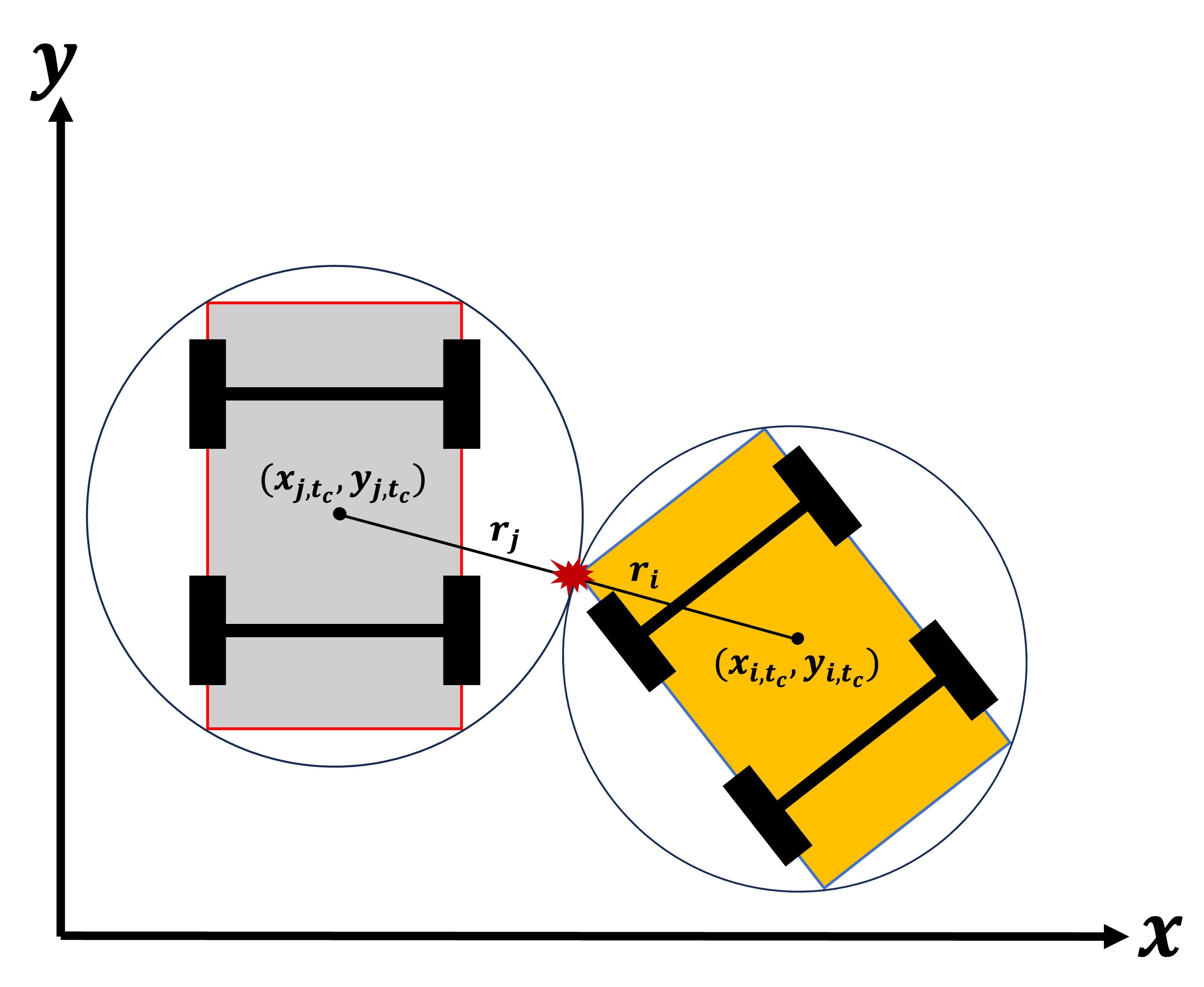}
    \caption{Description of Eqn.9}
    \label{fig:3}
\end{figure}

\bigskip
A mathematical condition to represent collisions was proposed by Li et al.(\citeyear{li2024beyond}), as shown in Eq. (\ref{eq5}). The equation calculates the euclidean distance between the centers of the bounding circles of two conflicting vehicles at coordinates $(x_i, y_i)$ and $(x_j, y_j)$. Setting the distance equal to the sum of their radii, $r_i + r_j$, implies that vehicles are deemed to collide with each other. Geometrically, this implies that the two bounding circles are tangential to each other (Fig:\ref{fig:3}), representing an imminent collision.

\begin{align}
    &g(\chi_i(t_c),\chi_{j}(t_c)) \nonumber\\
    =&(x_{i,t_c}-x_{j,t_c})^2+(y_{i,t_c}+y_{j,t_c})^2-(r_i+r_j)^2\nonumber\\
=&\biggl(-\frac{1}{12} \left(a_{i,0} v_{i,0} \sin(\theta_{i,0}) \frac{\tan(\delta_{i,0})}{L_i} - a_{j,0} v_{j,0} \sin(\theta_{j,0}) \frac{\tan(\delta_{j,0})}{L_j} \right) t_c^3 + \frac{1}{2} \biggl(a_{i,0} \cos(\theta_{i,0})-v_{i,0}^2\sin(\theta_{i,0})\frac{\tan(\delta_{i,0})}{L_i} \nonumber\\
&- a_{j,0} \cos(\theta_{j,0})+ v_{j,0}^2\sin(\theta_{j,0})\frac{\tan(\delta_{j,0})}{L_j}\biggr) t_c^2 + \left(v_{i,0}\cos(\theta_{i,0})-v_{j,0}\cos(\theta_{j,0})\right)t_c +(x_{i,0} - x_{j,0})\biggr)^2 \nonumber \\
&+ \biggl(\frac{1}{12} \biggl(a_{i,0} v_{i,0} \cos(\theta_{i,0}) \frac{\tan(\delta_{i,0})}{L_i} - a_{j,0} v_{j,0} \cos(\theta_{j,0}) \frac{\tan(\delta_{j,0})}{L_j} \biggr) t_c^3 \nonumber+ \frac{1}{2} \biggl(a_{i,0} \sin(\theta_{i,0}) +v_{i,0}^2\cos(\theta_{i,0})\frac{\tan(\delta_{i,0})}{L_i} \nonumber \\
&- a_{j,0} \sin(\theta_{j,0})-v_{j,0}^2\cos(\theta_{j,0})\frac{\tan(\delta_{j,0})}{L_j}\biggr) t_c^2 + \left(v_{i,0}\sin(\theta_{i,0})-v_{j,0}\sin(\theta_{j,0})\right)t_c+ (y_{i,0} - y_{j,0})\biggr)^2 - (r_i + r_j)^2  \label{eq5}
\end{align}

Here, $t_c$ is the time after which two vehicles would collide if they move with the same steering angle and acceleration. Subscript $i$ and $j$ denote the indices corresponding to individual vehicles, and $r$ is the radius of the vehicle from the center.

By setting \( g(x_i(t_c), x_j(t_c)) = 0 \) and solving for feasible \( t_c \), a novel 2D TTC is obtained, as detailed in Eq. (\ref{eq6}) and Fig. (\ref{fig:3}).

\begin{equation}
\text{TTC}_{ij} =
\begin{cases}
t_c, & \text{if } g(x_i(t_c), x_j(t_c)) = 0 \text{ and } t_c\geq0\\
\infty, & \text{otherwise}
\end{cases}
\label{eq6}
\end{equation}

\bigskip
AV sensor data contains granular information for vehicles \(i\) and \(j\) in each frame of \(z\). To identify the minimum TTC for any given pair of conflicting vehicles, we calculate the TTC between vehicle \(i\) and vehicle \(j\) for frame \(z\) denoted as \(\text{TTC}_{ij}(z)\). The following equation can determine the minimum TTC for any conflicting vehicle pair across all frames.

\begin{equation}
\text{X}=\text{TTC}_{\text{min}}(i,j) = \min_{z \in \{1, 2, \ldots, n\}} \{ \text{TTC}_{ij}(z) \}
\label{eq:ttcmin}
\end{equation}

This equation defines 
\text{X} as the minimum TTC value (near-miss events) observed between vehicles 
\(i\) and \(j\) over n frames for short duration segment.

\subsection{Univariate generalized extreme value (UGEV) distribution \label{sec3.3}}

In real-time crash estimation, EVT combined with near-miss events (observed values) provides a robust theoretical framework to infer latent crash occurrences by focusing on the distribution of rare, tail-end near-miss events (\citep{zheng2020novel}). Two primary approaches are employed for EVT modeling: Block maxima/minima (BM) and peak over threshold (POT). The BM approach divides the observation period into equal intervals, blocks, segments, and space, making it suitable for real-time frameworks because it systematically considers temporal variations (\citep{songchitruksa2006extreme}).  This study focuses on developing a real-time risk estimation framework; the BM approach is adopted for EVT modeling. That enables a structured and effective method for capturing and analyzing rare, high-risk events within the real-time context of traffic safety

This approach treats each interval's maximum or minimum value as an extreme event (\citep{coles2001introduction}). Including maxima or minima in the GEV model depends on the study's objective. They are assuming a common distribution for a series of independent random variables $X_1, X_2, \ldots, X_n$, within a block, the maximum values such as $M_n = \max \{-X_1, -X_2, \ldots, -X_n\}$ of $n$ observations, where X are derived from Eqn. \ref{eq:ttcmin} and used negative values for block maxima distribution. A GEV distribution will be converged if normalized maximum values $M_n^* = \frac{M_n - a_n}{b_n}$. 

\bigskip
In the univariate case, if there are sequences of constants $a_n > 0$ and $b_n$ such that: \[
\Pr\left(\frac{M_n - a_n}{b_n} \le X\right) \to F(X) \quad \text{as} \quad n \to \infty
\]

where $F$ is a non-degenerate distribution function that belongs to the Gumbel, Fréchet, or Weibull distribution families, which are all part of the GEV distribution (\citep{coles2001introduction}), shown in Eqn (\ref{eq7}).

\begin{equation}
f(X; \mu, \sigma, \xi) = \exp \left\{ -\left[1 + \xi \left(\frac{X - \mu}{\sigma}\right)\right]^{-1/\xi} \right\} \label{eq7}
\end{equation}

where $[1 + \xi \left(\frac{X - \mu}{\sigma}\right) > 0]$, $-\infty < \mu < \infty$, $\sigma > 0$, $-\infty < \xi < \infty$, $\xi > 0$ for Fréchet and $\xi < 0$ for Weibull, $\xi = 0$ for Gumbel distribution of GEV

\subsubsection{Hierarchical Bayesian structure random parameter
(HBSRP) model \label{sec3.3.1}}

This study employs a hierarchical Bayesian structure with a random parameter (HBSRP) model (\citep{zheng2020novel}), providing greater flexibility than fixed parameter models by accounting for unobserved heterogeneity across different sites. Through a multi-site strategy, the HBSRP model enhances effectiveness by using precise safety performance metrics from stable regions to inform less stable regions.  The Bayesian structure consists of three layers: data, process, and prior. Each block's extreme (TTC) values are modeled at the data level using a GEV distribution. The process level incorporates a Gaussian Process (GP) latent variable to capture spatial or temporal correlations and unobserved heterogeneity. The prior level involves selecting appropriate prior distributions, including their means and variances, to characterize each parameter and address uncertainties effectively.

By applying Bayes' theorem, the parameters $\Omega$ for the univariate extreme value model with a three-layer hierarchical structure given the data $X$, are inferred through the posterior distribution as formulated in Eqn (\ref{eq8}):

\begin{equation}
q(\Omega\mid X) \propto q_{\text{data}}(X\mid \Omega_1) q_{\text{process}}(\Omega_1 \mid \Omega_2) q_{\text{prior}}(\Omega_2) \label{eq8}
\end{equation}

Here, $q(\Omega\mid X)$ represents the posterior distribution of the parameters $\Omega$ given the observed data $X$. This posterior distribution is obtained by combining the likelihood of the data layer $q_{\text{data}}(X\mid \Omega_1)$, with the process layer $q_{\text{process}}(\Omega_1\mid \Omega_2)$ and prior layer $q_{\text{prior}}(\Omega_2)$. The density functions of $q_{\text{data}}, q_{\text{process}}, q_{\text{prior}}$ are given below.

\bigskip

The HBSRP model is formulated by denoting $k_n$ as the number of extreme traffic conflicts observed in the $n$-th block ($n=1,2,3,\ldots,k_n$) and $X_{nm}$ as the traffic conflict of the $n$-th block within the $m$-th site ($m=1,2,3,\ldots,i_m$). The magnitudes of the $k_n$ events occurring within a block are assumed to be realizations of independent and identically distributed random variables $X_{nm}$ with a common parametric cumulative distribution function $f$.

\bigskip
Due to the positive property of the scale parameter, the GEV distribution $(\mu, \sigma, \xi)$ is reparametrized as GEV $(\mu, \vartheta, \xi)$ where $(\vartheta = \log \sigma)$. Therefore, letting $\mu_m, \vartheta_m, \xi_m$ denote the parameters at the site $m$, the expression for the joint probability density function (pdf) of the reparameterized GEV distribution for all events is given by Eqn (\ref{eq9}):

\begin{equation}
q_{\text{data}}(X \mid \Omega_1) = \prod_{m=1}^{i} \prod_{n=1}^{k_m} \left( \frac{1}{\exp(\vartheta_m)} \right) \exp \left\{ - \left[1 + \xi_m \left(\frac{X_{nm} - \mu_m}{\exp(\vartheta_m)}\right)\right]^{-1/\xi_m} \right\} \left[1 + \xi_m \left(\frac{X_{nm} - \mu_m}{\exp(\vartheta_m)}\right)\right]^{-1-1/\xi_m} \label{eq9}
\end{equation}

The cumulative distribution function for $X_{nm}$ given the parameters $\mu_m, \vartheta_m, \xi_m$ is given by Eqn (\ref{eq10}):

\begin{equation}
f(X_{nm} < X \mid \mu_m, \sigma_m, \xi_m) = \exp \left\{ - \left[1 + \xi_m \left(\frac{X - \mu_m}{\exp(\vartheta_m)}\right)\right]^{-1/\xi_m} \right\} \label{eq10}
\end{equation}

This study utilizes a latent Gaussian process to model extremes, specifically linking data layer parameters to covariates through identity link functions. A significant challenge in this approach is accurately estimating the GEV distribution's shape parameter $\xi$. As it is unrealistic to construct a smooth function of covariates for $\xi$, it is generally treated as unknown and not influenced by covariates (\citep{coles2001introduction, cooley2006bayesian}). The model introduces random coefficients to address unobserved heterogeneity among observation sites, as shown in Eqn (\ref{eq11}). Random intercept coefficients capture site-specific heterogeneity unrelated to explanatory variables, while random coefficients account for the variation in covariate effects across different sites. This approach allows for a more thorough representation of unobserved heterogeneity, improving traffic safety analysis's accuracy and reliability by incorporating consistent site-specific factors and variable covariate impacts across sites.

\begin{equation}
\begin{aligned}
\mu_m &= \alpha_{\mu_{0m}} + \alpha_{\mu_m} Y \\
\vartheta_m &= \alpha_{\vartheta_{0m}} + \alpha_{\vartheta_m} Y \\
\xi_m &= \alpha_{\xi_{0m}} 
\end{aligned}
\label{eq11}
\end{equation}

Where, $\alpha_{\mu_{0m}}, \alpha_{\mu_m}, \alpha_{\vartheta_{0m}}, \alpha_{\vartheta_m}, \alpha_{\xi_{0m}}$ are random coefficients and $Y$ is an explanatory variable (i.e., volume, speed, acceleration, dimensions, gaps, etc.). With these parameters, the process layer works according to Eqn (\ref{eq12}):

\begin{equation}
q_{\text{process}}(\Omega_1 \mid \Omega_2) = \frac{1}{\sqrt{2\pi \tau_\mu^2}} \exp\left\{ -\frac{1}{2\tau_\mu^2} (\mu - \mu_m)^2 \right\} \times \frac{1}{\sqrt{2\pi \tau_\vartheta^2}} \exp\left\{ -\frac{1}{2\tau_\vartheta^2} (\vartheta - \vartheta_m)^2 \right\} \times \frac{1}{\sqrt{2\pi \tau_\xi^2}} \exp\left\{ -\frac{1}{2\tau_\xi^2} (\xi - \xi_m)^2 \right\}
\label{eq12}
\end{equation}

In Bayesian analysis, it is crucial to identify appropriate prior distributions. For this study, noninformative priors were selected empirically. The model parameters $\alpha_{\mu_{0m}}, \alpha_{\mu_m}, \alpha_{\vartheta_{0m}}, \alpha_{\vartheta_m}, \alpha_{\xi_{0m}}$ are assigned normal prior distributions, specifically $N(\delta_{\mu}, \delta_{\sigma})$. According to Fu et al., (\citeyear{fu2020multivariate}), the mean $\delta_{\mu}$ of these distributions is itself normally distributed $\delta_{\mu} \sim N(0, 0.00001)$. The variance $\delta_{\sigma}$ follows an inverse gamma distribution $\delta_{\sigma} \sim \text{IG}(0.001, 0.001)$. The GEV distribution's shape parameter $\xi$ is typically uniformly distributed between -1 and 1. The criterion for selecting these prior distributions is the successful convergence of the model, and the prior works as Equation (\ref{eq13}).

\begin{equation}
q_{\text{prior}}(\Omega_2) = q_{\alpha_{\mu_{0m}}} (\alpha_{\mu_{0m}}) \times q_{\alpha_{\mu_m}} (\alpha_m) \times q_{\alpha_{\vartheta_{0m}}} (\alpha_{\vartheta_{0m}}) \times q_{\alpha_{\vartheta_m}} (\alpha_m) \times q_{\alpha_{\xi_{0m}}} (\alpha_{\xi_{0m}}) \label{eq13}
\end{equation}

\bigskip
\subsubsection{Model choice \label{sec3.3.2}}

Using the HBSRP model framework, several models can be developed by reparameterizing location and scale parameters in the GEV distribution and introducing different combinations of covariates. To identify the most parsimonious model among these models, we compare them using the Deviance Information Criterion (DIC), a well-established metric for Bayesian model selection. The principle of DIC is parsimony, aiming to find the simplest model that explains the most variation in the data (\citep{spiegelhalter2002bayesian}). DIC balances model complexity and goodness of fit, effectively preventing overfitting in hierarchical models while enabling straightforward comparisons by integrating fit and complexity into one metric. The DIC is calculated as Eqn (\ref{eq14}):

\begin{equation}
\text{DIC} = \bar{D} + p_D \label{eq14}
\end{equation}

where $\bar{D}$ is the posterior mean deviance, indicating model fit, and $p_D$ is the adequate number of parameters. Generally, a model with a lower DIC is preferred. A difference greater than 10 in DIC values between models strongly favors the model with the lower DIC. Differences between 5 and 10 are considered substantial, while differences less than 5 suggest that the models are competitive (\citep{el2012measuringa,el2012measuringb}).

\subsubsection{Crash estimation}\label{sec3.3.3}

The Risk of Crash (RC) index is a probabilistic measure derived from the GEV distribution, used to quantify the likelihood of a crash based on TTC data. In this context, a crash is defined as having a TTC equal to zero. The RC index represents the probability that the maximum negated TTC in a given block exceeds zero. The likelihood for block $n$ is given by Eqn (\ref{eq15}):

\begin{equation}
\text{RC}_n = \Pr(X_n \geq 0) =
\begin{cases}
1 - \exp \left\{ -\left[1 - \xi_n \left(\frac{\mu_n}{\exp(\vartheta_n)}\right)\right]^{-1/\xi_n} \right\}, & \xi_n \neq 0 \\
1 - \exp \left\{ -\exp \left(\frac{\mu_n}{\exp(\vartheta_n)}\right) \right\}, & \xi_n = 0
\end{cases}
\label{eq15}
\end{equation}

In this equation, $\xi_n$, $\vartheta_n$, and $\mu_n$ are the GEV distribution's shape, scale, and location parameters, respectively. The value of $\text{RC}_n$ ranges from 0 to 1, where 0 indicates no crash risk and values closer to 1 indicate higher crash risk. This index provides a quantitative measure of crash probability, enabling researchers and engineers to identify high-risk areas and implement safety interventions accordingly. 

\bigskip

The estimated crash risk can be translated to extended periods beyond the observation period, such as annually. To calculate the crash frequency for both the observation period and yearly, use Eqn (\ref{eq16}):

\begin{equation}
\text{CF} = \sum_{n=1}^{k} \text{RC}_n 
\label{eq16}
\end{equation}

\begin{equation}
\text{CF}_{\text{year}} = \frac{T}{t} \times \text{CF}
\label{eq17}
\end{equation}

where CF is defined as crash frequency, $t$ is the number of BM used in the observation period, and $T$ is the total annual BM data in the model.

\section{Analysis and results}\label{sec4}

\subsection{Data preparation \label{sec4.1}}
\begin{figure}[h!]
    \centering
    \setlength{\abovecaptionskip}{0pt}
    \subcaptionbox{Site-1}
    {\includegraphics[width=0.4\textwidth]{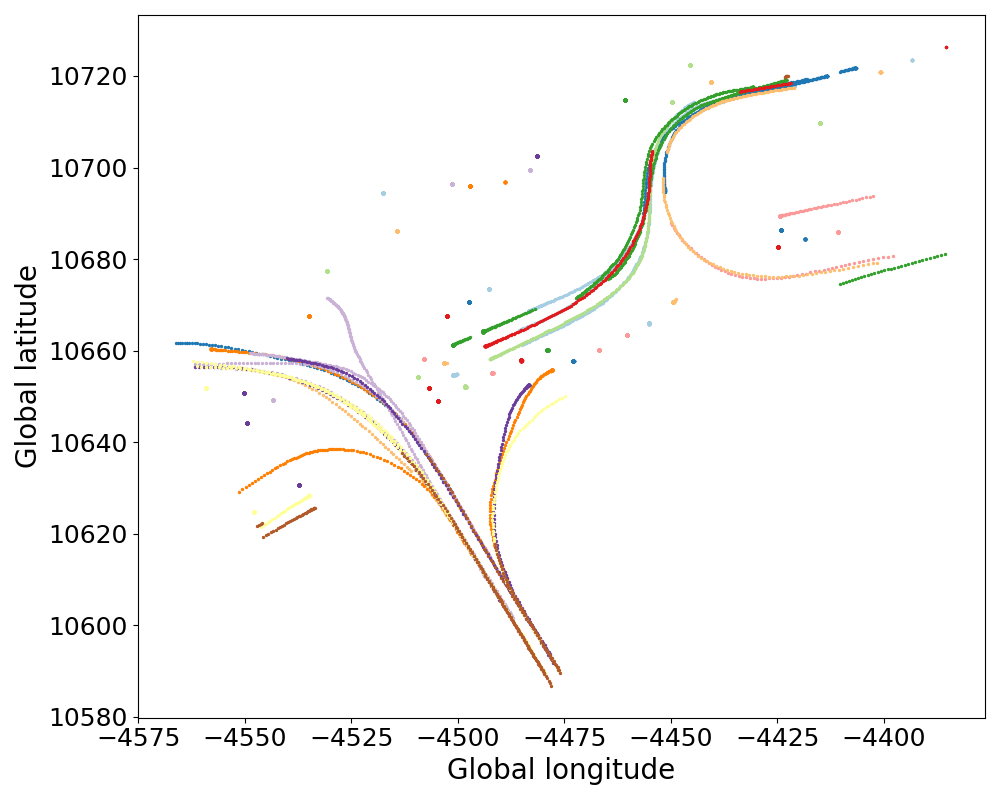}}
    \subcaptionbox{Site-2}{\includegraphics[width=0.4\textwidth]{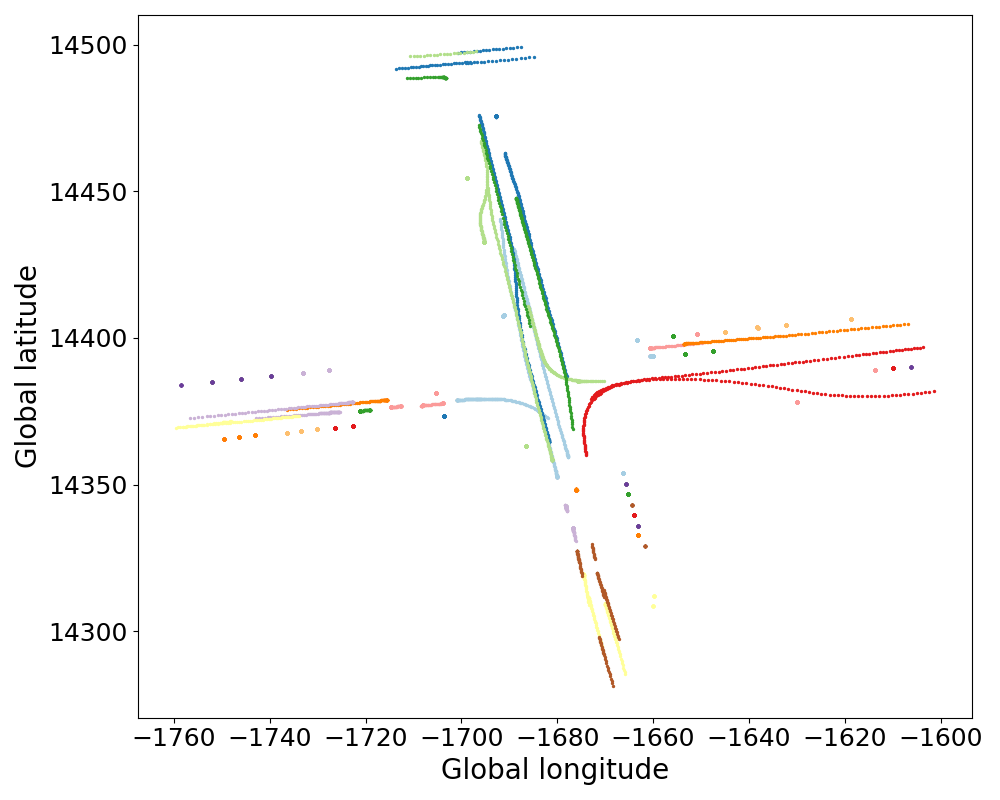}}
    \subcaptionbox{Site-3}{\includegraphics[width=0.4\textwidth]{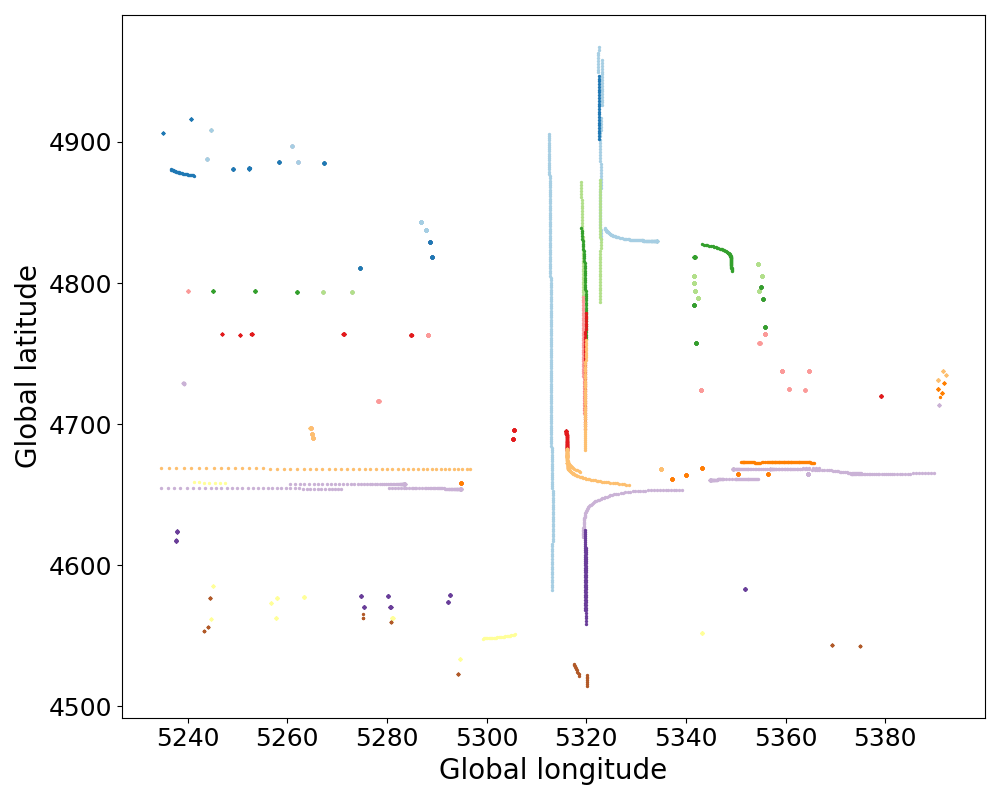}}
    \subcaptionbox{Site-4}{\includegraphics[width=0.4\textwidth]{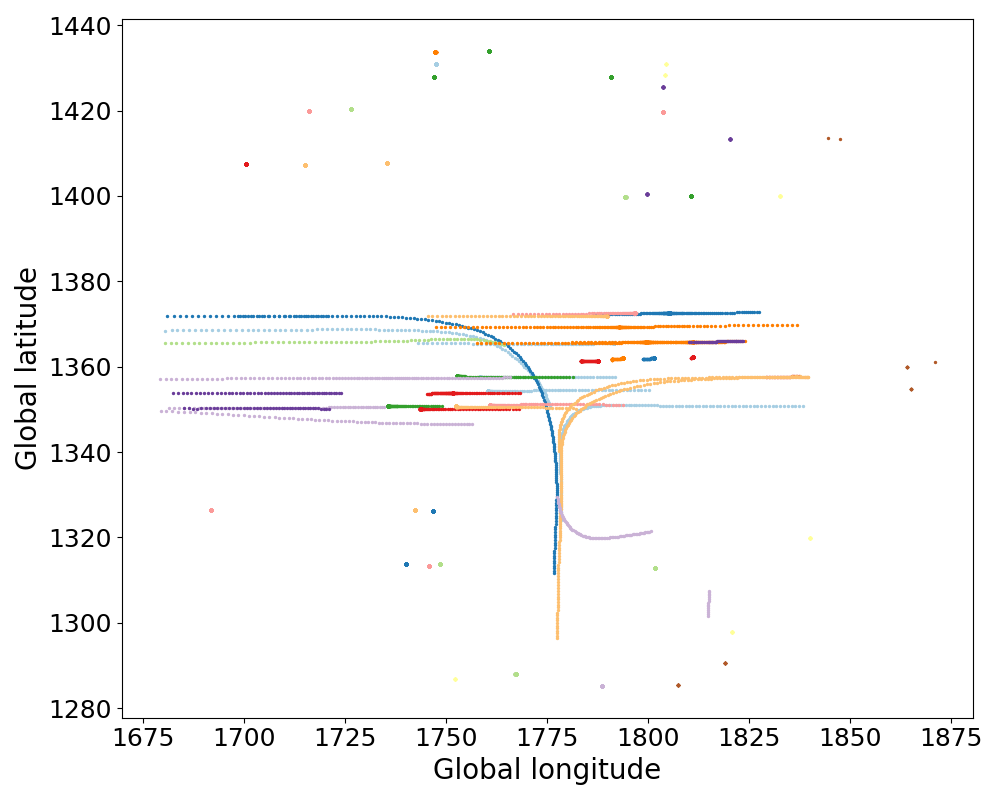}}
    \subcaptionbox{Site-5}{\includegraphics[width=0.4\textwidth]{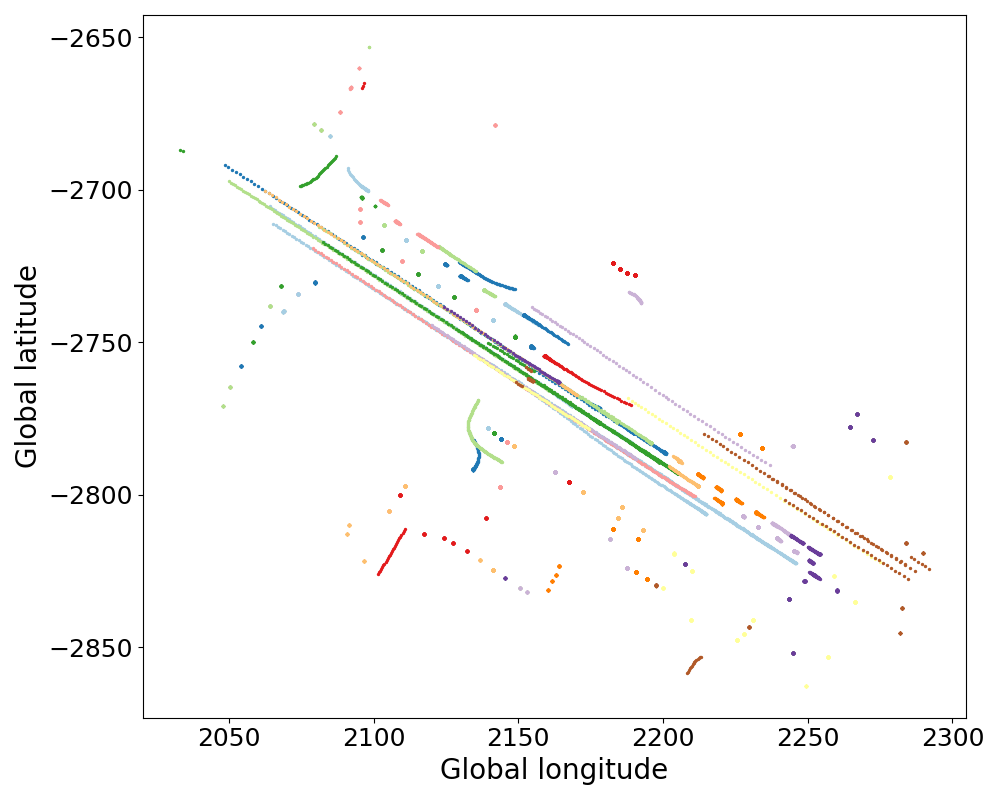}}
    \subcaptionbox{Site-6}{\includegraphics[width=0.4\textwidth]{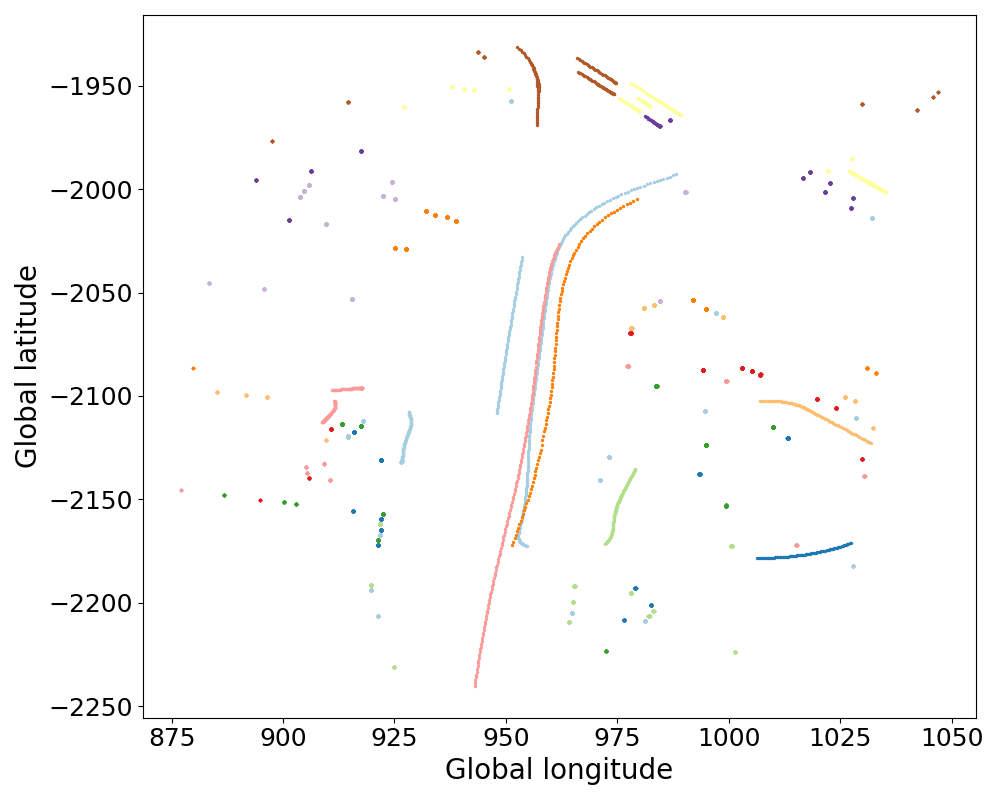}}    
    \caption{The study locations: San Francisco (1\&2), Phonix (3\&4), Los Angeles (5\&6)}
    \label{fig:4}
\end{figure}

The dataset for this study was meticulously extracted from the Waymo fleet data (described in section \ref{sec3.1}), currently operating in various urban networks across cities such as San Francisco, Phoenix, and Los Angeles. Two segments were selected ( see Fig. \ref{fig:4}) for each of these cities based on specific criteria designed to capture diverse and complex traffic scenarios as depicted in Fig. \ref{fig:2}. The selected segments considered high traffic movement, vehicle size variability, lane-changing vehicles, midblock sections, and intersections to represent diverse driving behaviors and capture more complex 2D movements. 

The data preparation for this analysis involved several key steps to ensure a targeted focus on vehicle-to-vehicle interactions. In the initial data extraction, all object types were filtered, retaining only data points classified as vehicles while excluding pedestrians and cyclists. This approach was chosen to analyze traffic interactions between vehicles (AV or HDV), which is the primary focus of this study. The exclusion of other road users (pedestrians or cyclists) was intended to reduce complexity and provide a more precise analysis of vehicle dynamics, thereby enhancing the robustness of the study’s findings regarding near-miss events. Additionally, extracted data was cleaned by removing anomalies, duplicates, or irrelevant information. This step ensured the dataset's integrity and accuracy, providing a reliable foundation for further analysis. The cleaned data was segmented for each city and all detected objects' information and other relevant information. The dataset contains vehicle information, traffic parameters, and other pertinent information as listed in Table \ref{Table:1}.  

The dataset includes real-time local positions of all detected traffic identified by the LiDAR system of the Waymo ego vehicle at each time step.  It is necessary to transform the HDVs' local (ego vehicle: Waymo) headings into global headings. While the Waymo vehicle’s heading ($\alpha$) is already in the global coordinate system, the HDVs’ headings ($\alpha'$) are measured in the local coordinate system relative to the ego vehicle. Therefore, to convert these local headings into the global coordinate system, we add the Waymo vehicle’s global heading to each HDV’s local heading. This transformation process, illustrated in Fig. \ref{fig:5}(a), is expressed mathematically as:

\begin{figure}[h]
    \centering
    \setlength{\abovecaptionskip}{0pt}
    \subcaptionbox{Heading}
    {\includegraphics[width=0.6\textwidth]{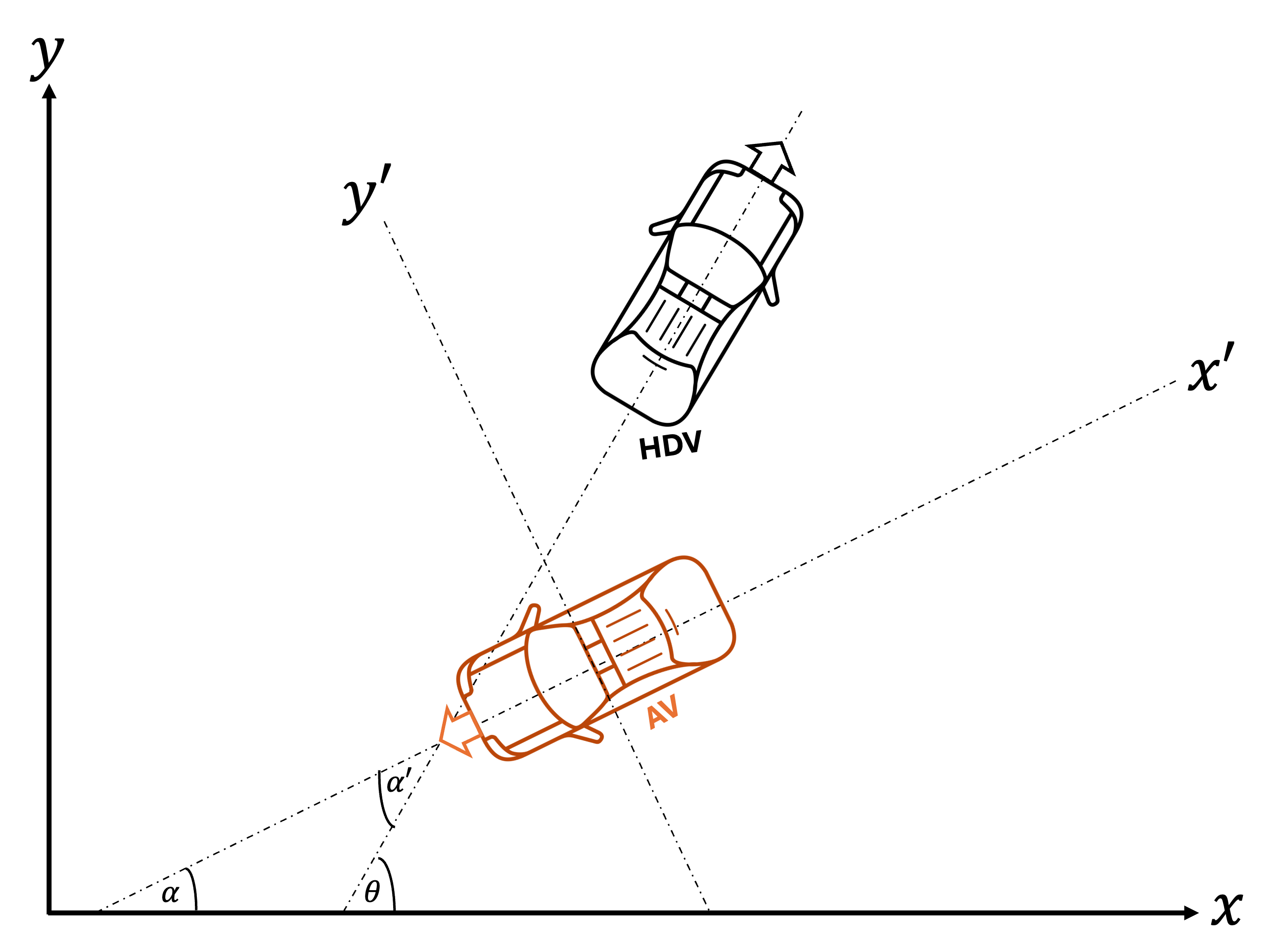}}
    \subcaptionbox{Speed}{\includegraphics[width=0.4\textwidth]{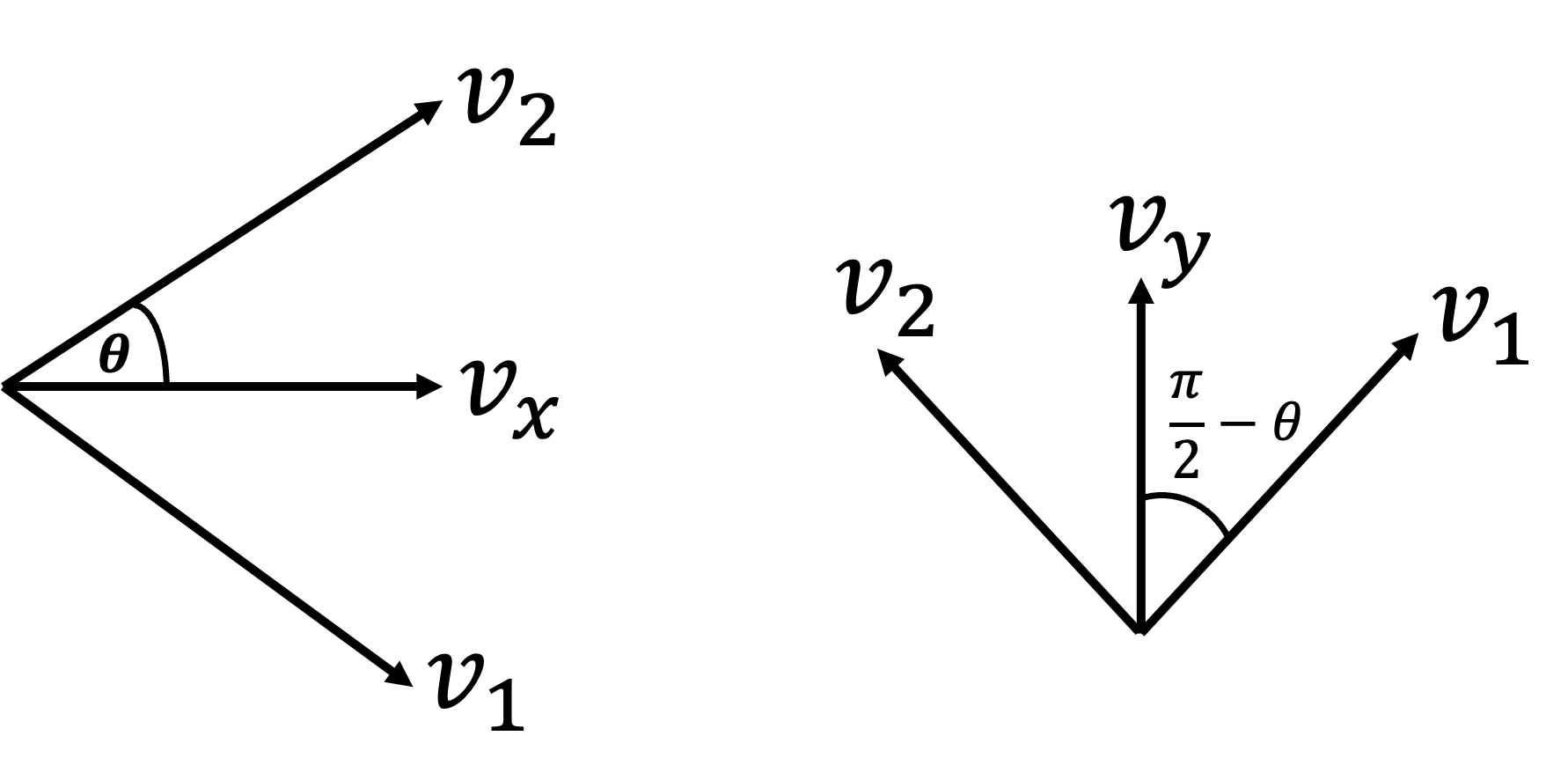}}
    \subcaptionbox{Acceleration}{\includegraphics[width=0.4\textwidth]{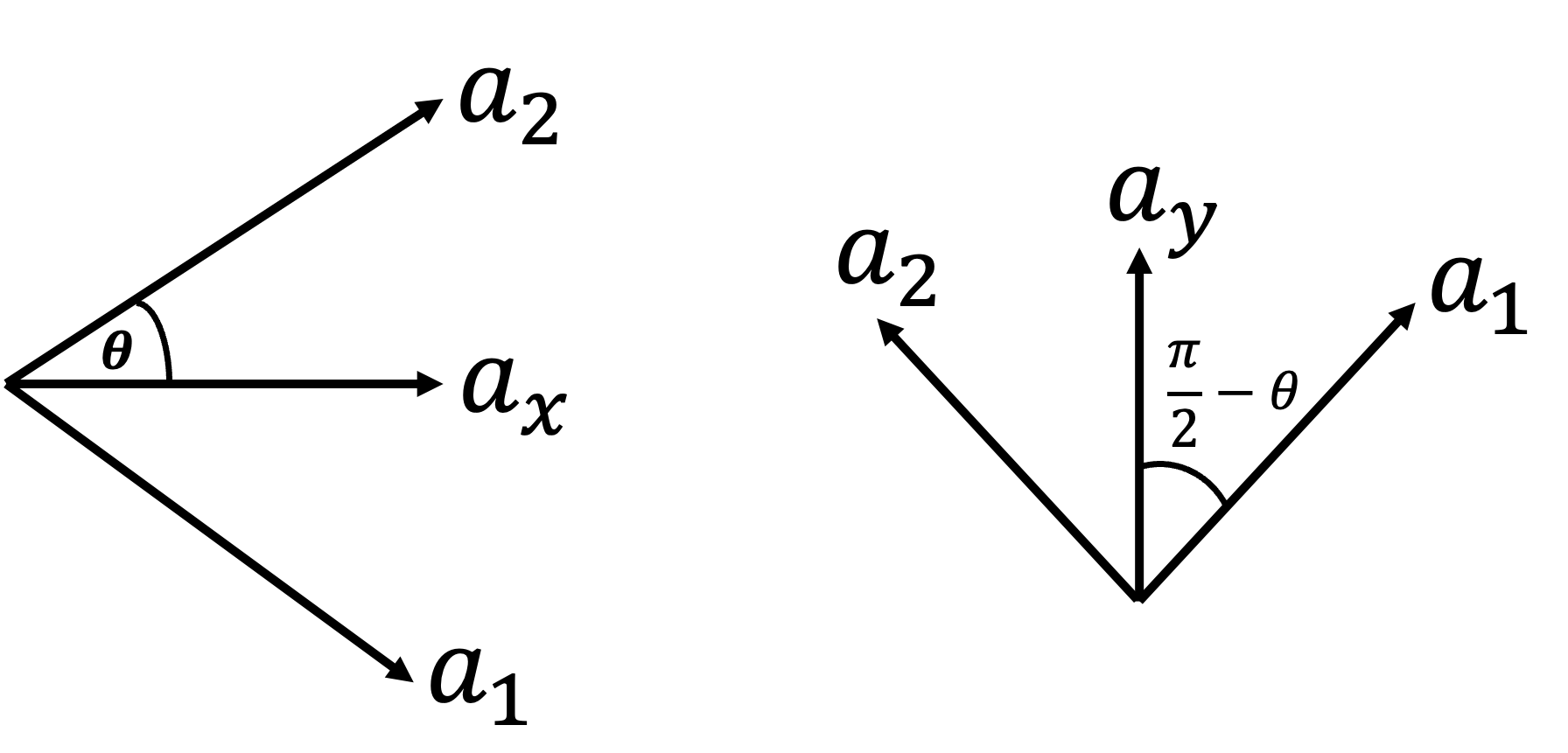}}
    \caption{Transformation to the global reference}
    \label{fig:5}
\end{figure}

\begin{equation}
\begin{aligned}
\theta &= \alpha + \alpha' 
\end{aligned}
\label{eq18}
\end{equation}

where $\theta$ represents the global heading of an HDV. This step is crucial for ensuring that the orientation of each HDV is accurately represented in the global context, which is essential for task trajectory calculation. Similarly, speed, acceleration, and steering angle data were transformed into the global reference frame for all HDV vehicles, as shown in Fig. \ref{fig:5}(b)\&(c). The transformations are given by:

\begin{equation}
\begin{aligned}
v &= v_x \cos \theta + v_y \cos\left(\frac{\pi}{2} - \theta\right) 
\end{aligned}
\label{eq19}
\end{equation}

\begin{equation}
\begin{aligned}
a &= a_x \cos \theta + a_y \cos\left(\frac{\pi}{2} - \theta\right) 
\end{aligned}
\label{eq20}
\end{equation}

\begin{equation}
\begin{aligned}
\delta &= \tan^{-1} \left(\frac{\dot{\theta} L}{v}\right)
\end{aligned}
\label{eq21}
\end{equation}

where $v$ is the global speed, $v_x$ and $v_y$ are the speed components in the local coordinate system, $a$ is the global acceleration, $a_x$ and $a_y$ are the acceleration components in the local coordinate system, $\delta$ is the global steering angle, $\dot{\theta}$ is the rate of change of the heading angle in each frame, and $L$ is the vehicle's wheelbase. The transformations are crucial to accurately represent the speed, acceleration, heading, and steering angle data of HDVs in the global coordinate system. Once these parameters are transformed into the global reference frame, summary statistics are prepared for each site, as detailed in Table \ref{table:2}. The approximate vehicle's radius ($r$) is derived from available object dimensions $L$ and $W$. To obtain the $t_c$ value between two vehicles, whether HDV or AV, the necessary information is applied using Eqn. \ref{eq5}. This study  $t_c$ < 3s is considered a near-miss event; previous studies set the threshold from 0.1s to 3s (\citep{fu2022random,fu2022bayesian,zhang2019real,kamel2023real,kamel2024real}).

\begin{table}[ht]
   \caption{Summary data statistics of six sites}
   \label{table:2}
   \centering
   \scriptsize  
   \renewcommand{\arraystretch}{0.8}
   \resizebox{\textwidth}{!}{
    \begin{tabular}{llcccccc}
    \hline
    Variable & Descriptive statistics & Site-1 & Site-2 & Site-3 & Site-4 & Site-5 & Site-6 \\
    \\
    \hline
    Volume  & Min &45 &22 &15 &43 & 61&41 \\
           & Max &71 & 69&48 &53 &83 &63 \\
           & Mean &62.90 &53.85 & 33.94& 46.54& 76.89&54.95 \\   
           & Std. Dev. &6.04 &13.52 &6.82 & 2.74& 3.31&5.32 \\  
    Speed  & Min & -0.18&-0.95 &-0.44 &-0.20 &-0.13 &-1.65 \\
          & Max &15.71 & 10.38& 22.20&15.55 & 19.96& 19.09\\
          & Mean &1.94 &1.22 &2.29 &1.68 &1.86 & 0.82\\   
          & Std. Dev. &3.16 & 2.38&5.37 &3.21 &3.86 &2.82 \\
    Acceleration  & Min &-5.0 & -4.97&-2.97 & -4.09&-6.47 &-2.69 \\
                 & Max &3.88 &3.50 &4.56 & 2.89&3.00 &3.01 \\
                 & Mean &-0.002 &-0.027 &-0.02 &0.99 &-0.16 &-0.03 \\   
                 & Std. Dev. & 0.54& 0.52&0.52 & 0.63& 0.54& 0.27\\         
    Heading  & Min &-1.57 &-1.37 &-2.2 & -1.32&-1.57 & -1.42\\
            & Max &1.57 &1.37 &2.2 & 1.32&1.57 &1.42 \\
            & Mean & 0.007& -0.007&-0.01 &0.04 & -.002&0.009 \\   
            & Std. Dev. &0.59 &0.72 & 0.76&0.45 & 0.80&0.94 \\  
    Steering angle  & Min &-5.85 & -4.56& -4.71& -3.15&-3.77 &-1.83 \\
                   & Max &0.77 & 1.73& 1.57& 3.14& 2.59& 4.78\\
                   & Mean & -2.94&-1.73 &-1.55 &-0.64 &-0.89 &1.61 \\   
                   & Std. Dev. &1.52 &1.43 &1.72 &1.88 &1.74 &1.70 \\  
    Length  & Min & 3.52&3.18 &3.5 & 3.95&3.24 & 3.0\\
           & Max & 12.65&14.28 &7.85 & 6.92&6.41 & 14.86\\
           & Mean & 4.79& 5.10& 5.10& 4.78&4.58 & 4.74\\   
           & Std. Dev. &1.11 &1.30 &0.87 &0.54 & 0.49&1.10 \\  
    Width  & Min & 1.76&1.81 &1.82 & 1.81& 1.36& 1.41\\
          & Max &3.33 &4 & 3.16& 2.92&2.95 & 3.08\\
          & Mean & 2.15&2.22 & 2.19& 2.12&2.04 &2.04 \\   
          & Std. Dev. &0.24 & 0.30&0.28 &0.18 &0.20 &0.23 \\    
    \hline
    \end{tabular}
   }
   \normalsize
\end{table}

\begin{figure}[h]
    \centering
    \setlength{\abovecaptionskip}{0pt}
    \includegraphics[width=0.8\textwidth]{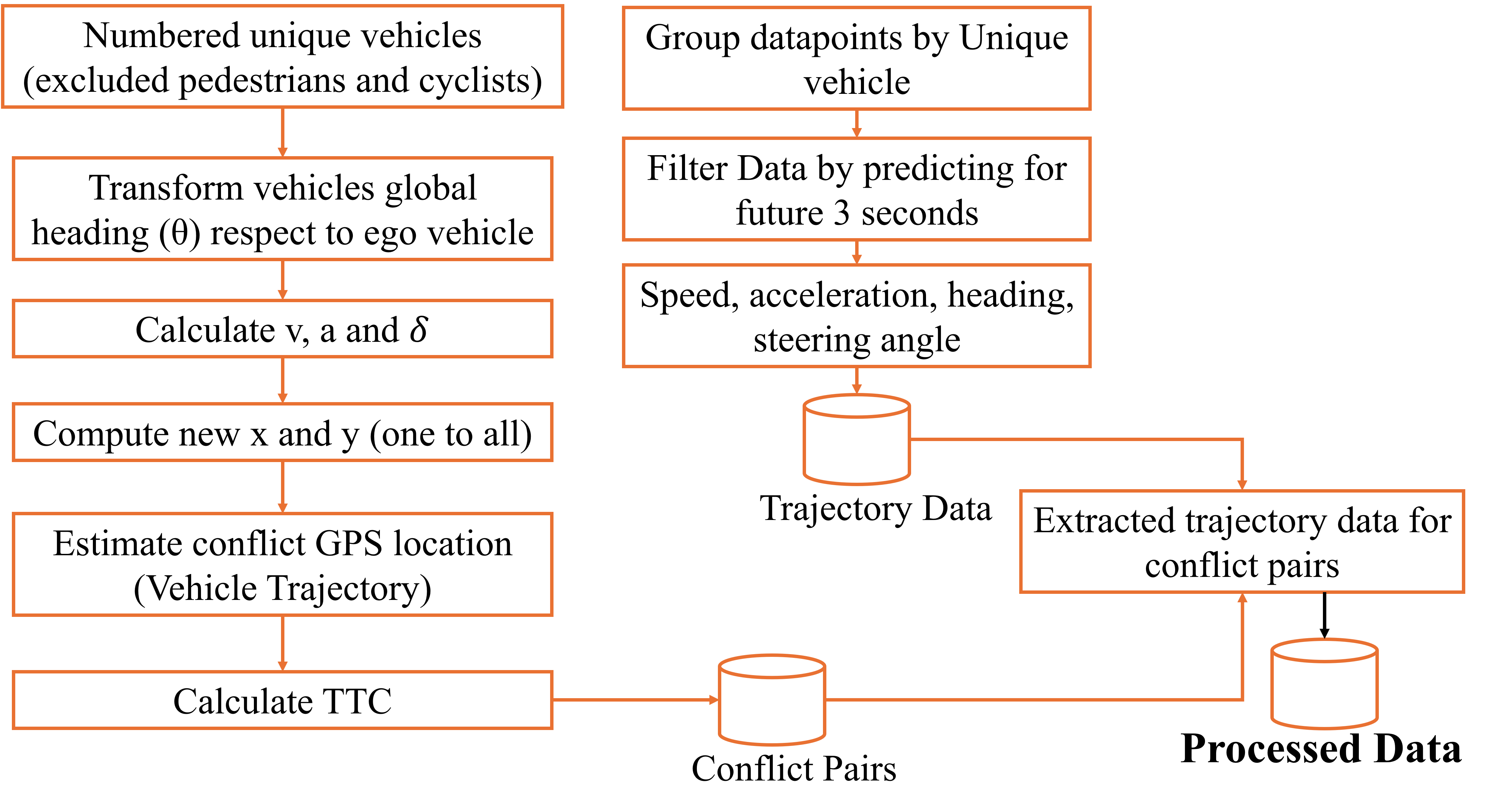}
    \caption{Data processing pipeline for calculating $TTC_{ij}$ from conflicting pairs}
    \label{fig:6}
\end{figure}

\begin{figure}[h!]
    \centering
    \setlength{\abovecaptionskip}{0pt}
    \subcaptionbox{Frame-1}
    {\includegraphics[width=0.3\textwidth]{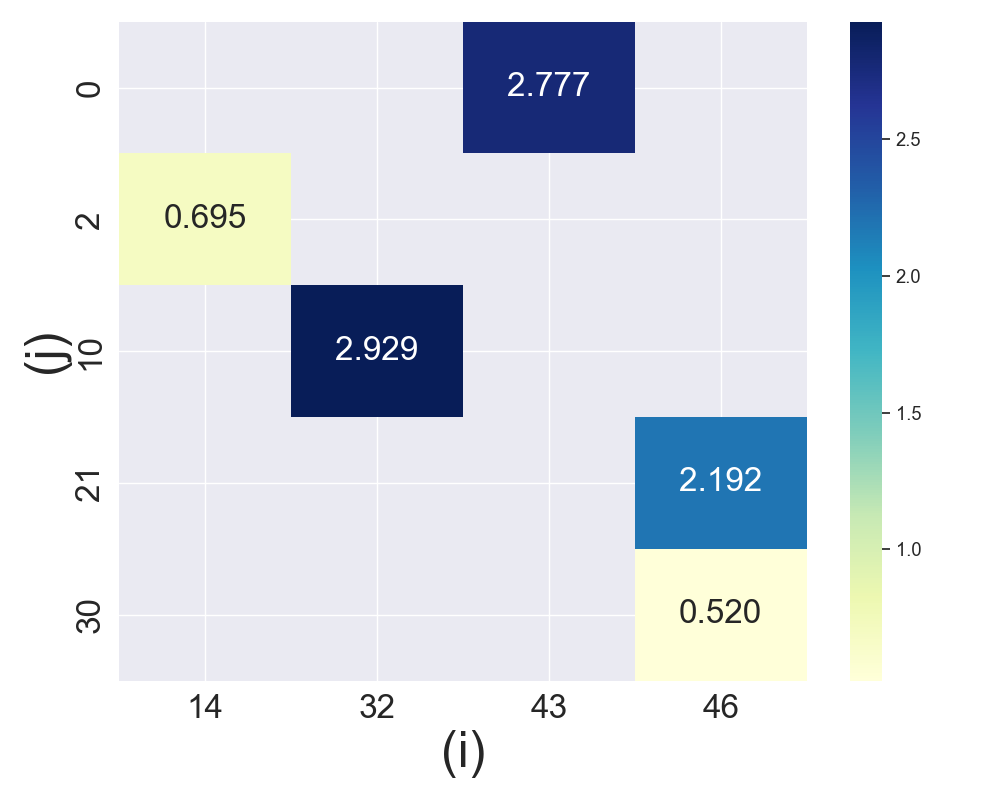}}
    \subcaptionbox{Frame-2}{\includegraphics[width=0.3\textwidth]{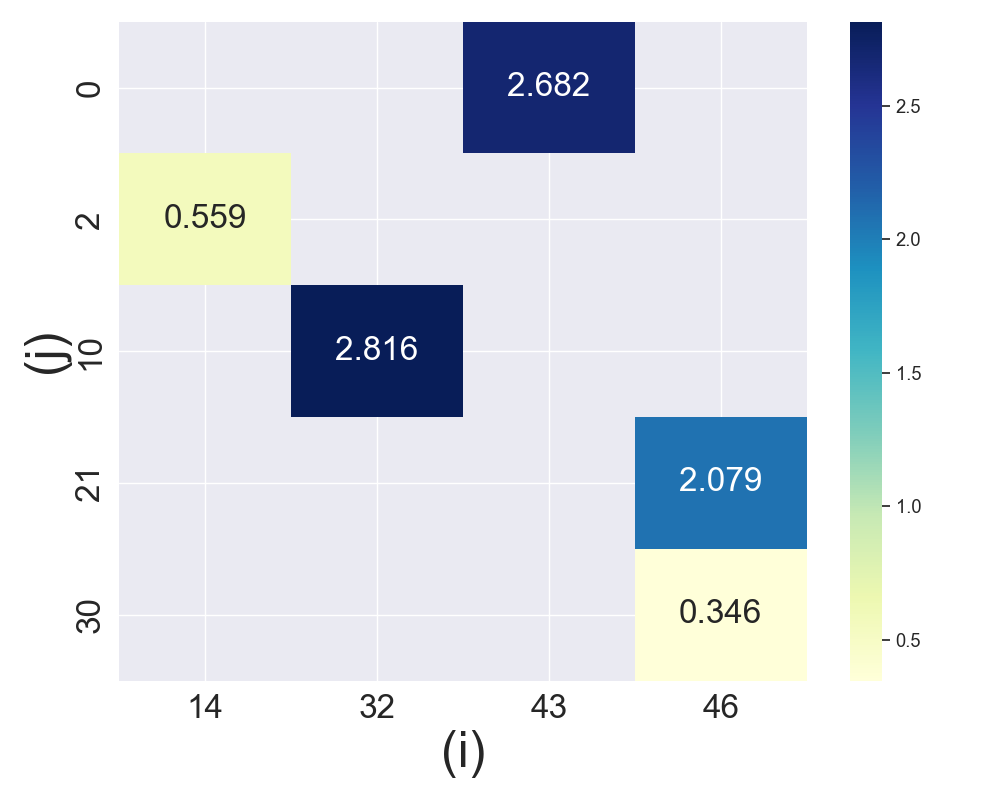}}
    \subcaptionbox{Frame-3}{\includegraphics[width=0.3\textwidth]{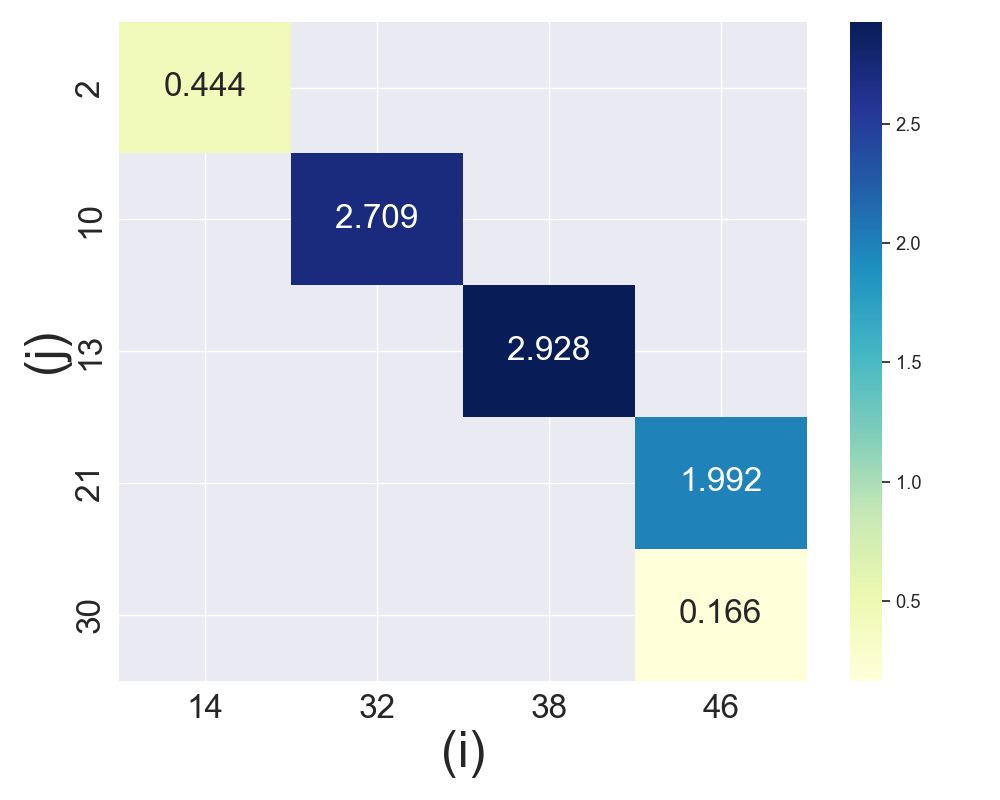}}
    \subcaptionbox{Frame-4}{\includegraphics[width=0.3\textwidth]{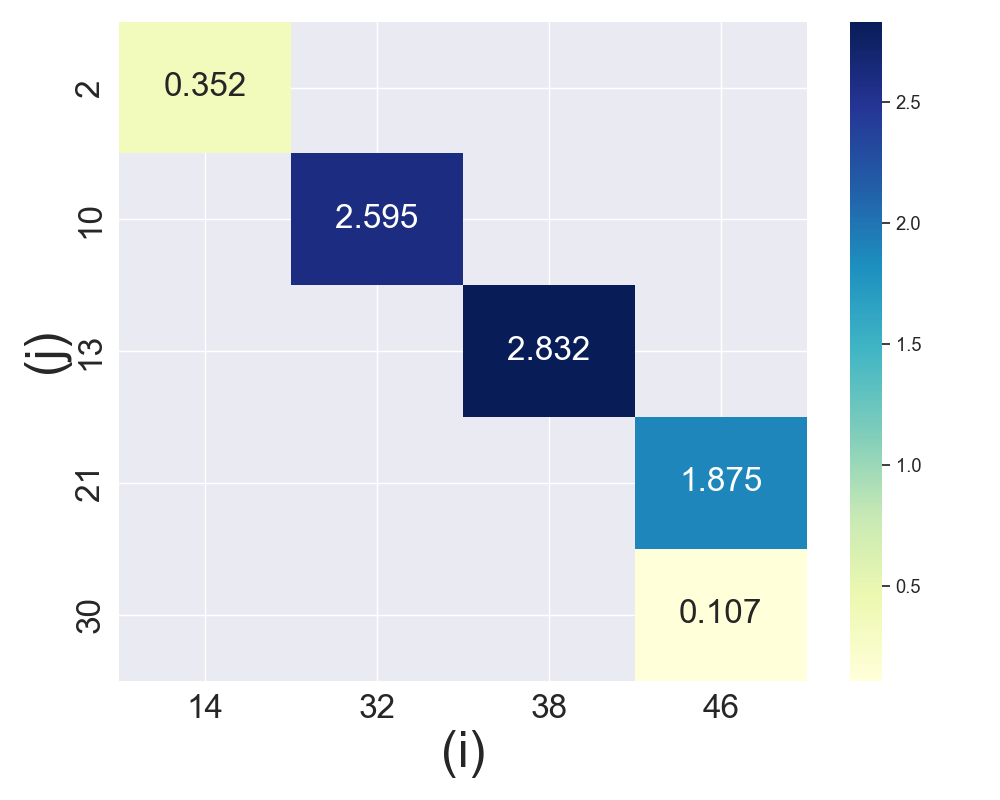}}
    \subcaptionbox{Frame-5}{\includegraphics[width=0.3\textwidth]{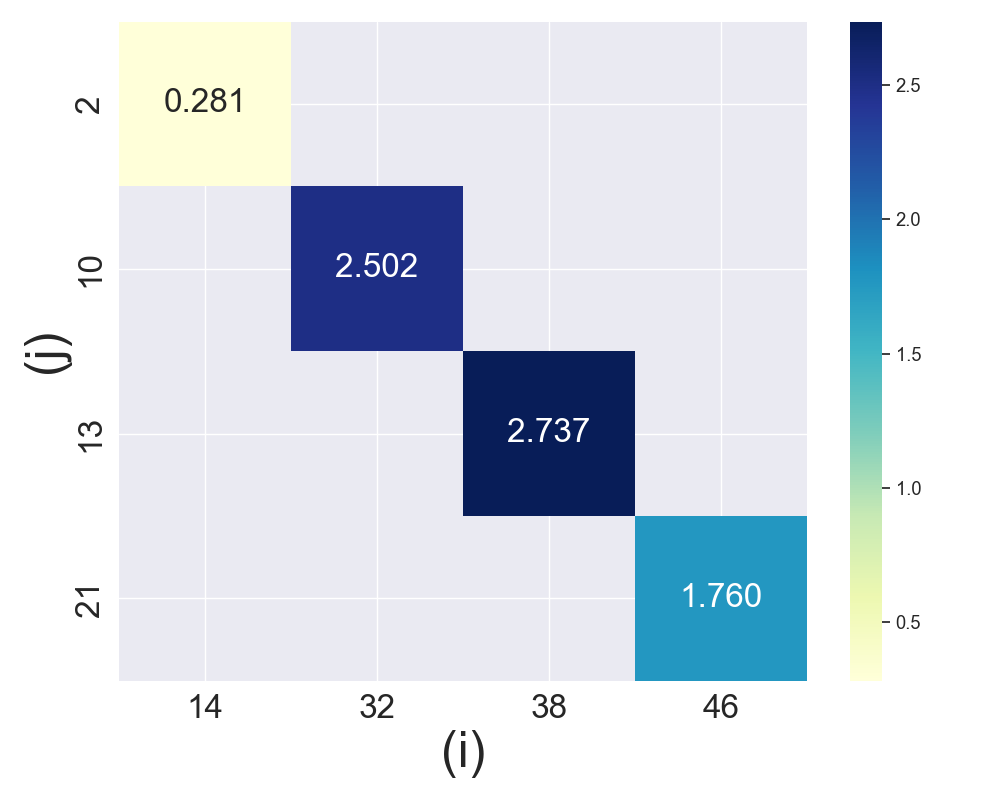}}
    \subcaptionbox{Frame-6}{\includegraphics[width=0.3\textwidth]{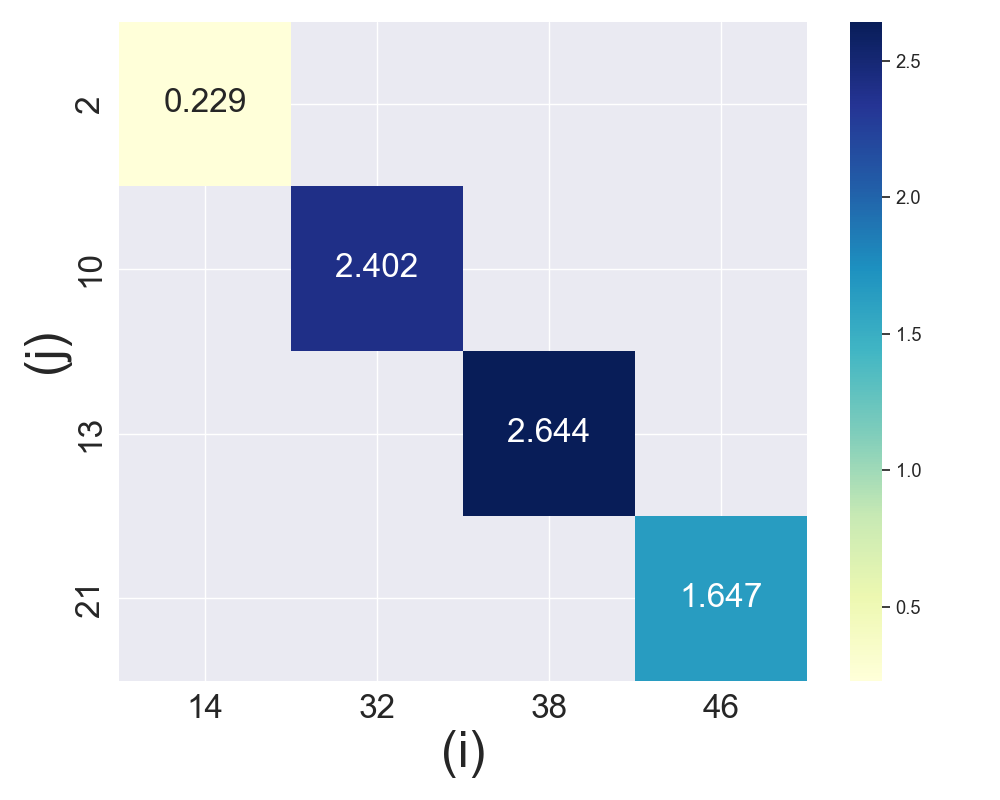}}    
    \caption{$TTC_{ij}$ calculation of conflicting pairs}
    \label{fig:7}
\end{figure}

The data processing pipeline, illustrated in Fig. \ref{fig:6}, employs two separate pipelines: one for identifying conflict pairs (according to Eqn.\ref{eq5}) and another for their processing trajectory data. This dual-pipeline approach facilitated the efficient handling and analysis of the complex dataset. To identify conflicting vehicle pairs, the method calculates the  2D TTC for all vehicles over 200 frames, a set boundary of 0.1 to 3s using Eqns (\ref{eq5}) and (\ref{eq6}) with parameters listed in Table \ref{table:2}. This frame-by-frame identification of conflict pairs is illustrated in Fig. \ref{fig:7} (a subset of 200 frames: $t_c$ and conflicting pairs).

\bigskip

The subset of sequential 2D $TTC_{ij}$ ($t_c$) calculations for distinct conflicting vehicle pairs throughout 200 frames are illustrated in Fig. \ref{fig:7}. These sequential frames capture the dynamic interactions between two vehicles, denoted as $i$ and $j$, focusing on the continuous reduction of the $t_c$ value as their proximity increases throughout the progression of frames. In Frame 1, the $t_c$ value is computed using Eqns \ref{eq5} and \ref{eq6}, incorporating the parameters $x$, $y$, $v$, $a$, $\theta$, $\delta$, $L$, and $r$, where $i \in V$ and $V$ represents the set of all vehicles (considering two vehicles indexed by $i$ and $j$ in each frame $z$). As the sequence proceeds ($z \in \{1, 2, \ldots, n\}$), the trajectories of vehicles $i$ and $j$ are updated, and their $t_c$ values are recalculated to reflect their relative positions and distances. Frames 2 through 6 depict intermediate states, wherein the $t_c$ values continue to decrease as vehicles $i$ and $j$ approach one another until a course change occurs. For example, vehicles 30 and 46 form a conflicting pair, with their $t_c$ values decreasing until Frame 4, beyond which they disappear, indicating a change in heading. In contrast, vehicles 21 and 46 are observed across all Frames 1 to 6, remaining near a minimum $t_c$ value of 1.647 seconds, qualifying as a conflicting pair under the study's criteria. Any vehicle pair that remains in conflict across all 200 frames ($z \in \{1, 2, \ldots, n\}$) is designated as a persistent conflicting pair, their min $t_c$ representing BM, as described in Eqn \ref{eq6}. It is important to note that Fig. \ref{fig:7} presents only a subset of conflicting pair's ($i$ and $j$) $t_c$ values. The computed $t_c$ values represent scenarios assuming that the involved vehicles take no evasive actions. This detailed sequence of frames demonstrates the real-time calculation of $t_c$, elucidating the dynamics leading up to a potential collision.

\bigskip

\begin{figure}[h!]
    \centering
    \setlength{\abovecaptionskip}{0pt}
    \subcaptionbox{Relationship among volume (vol
), speed (v), acceleration (a), and TTC}
    {\includegraphics[width=0.8\textwidth]{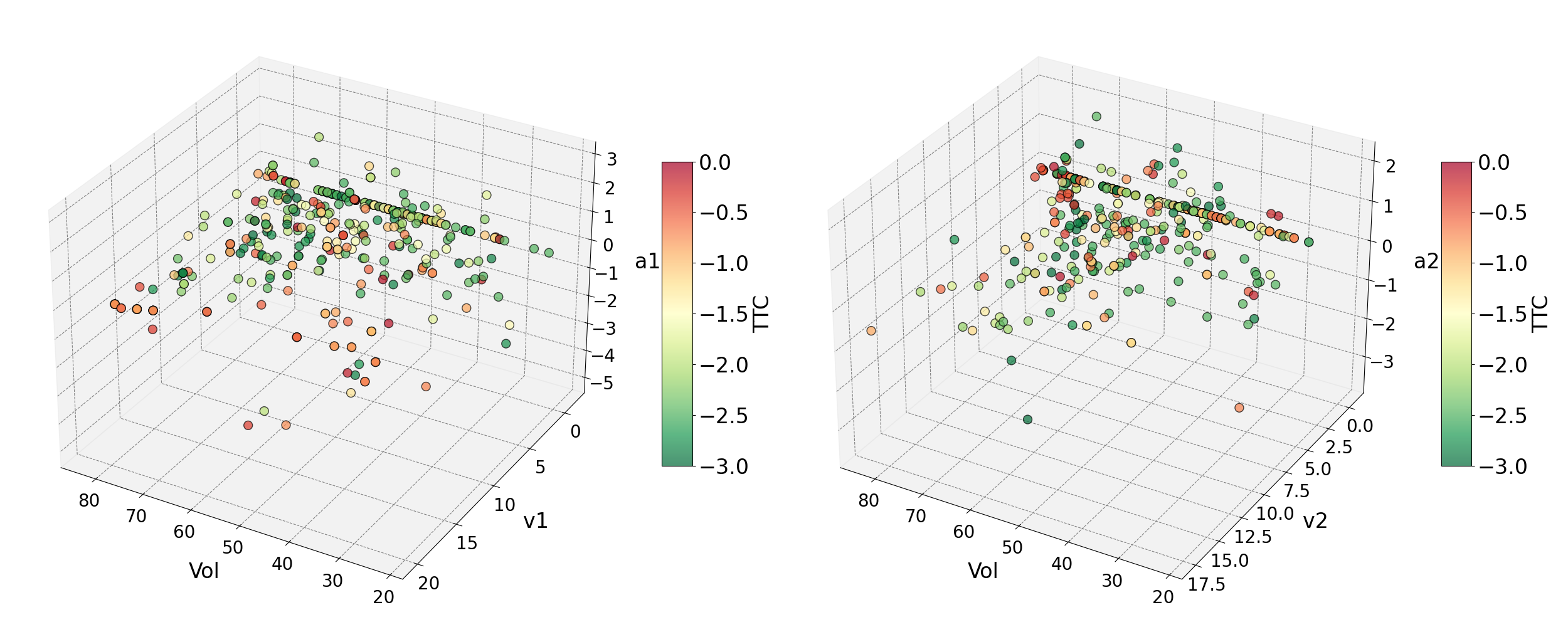}}
    \subcaptionbox{Relationship among length (L), speed (v), acceleration (a) and TTC}{\includegraphics[width=0.8\textwidth]{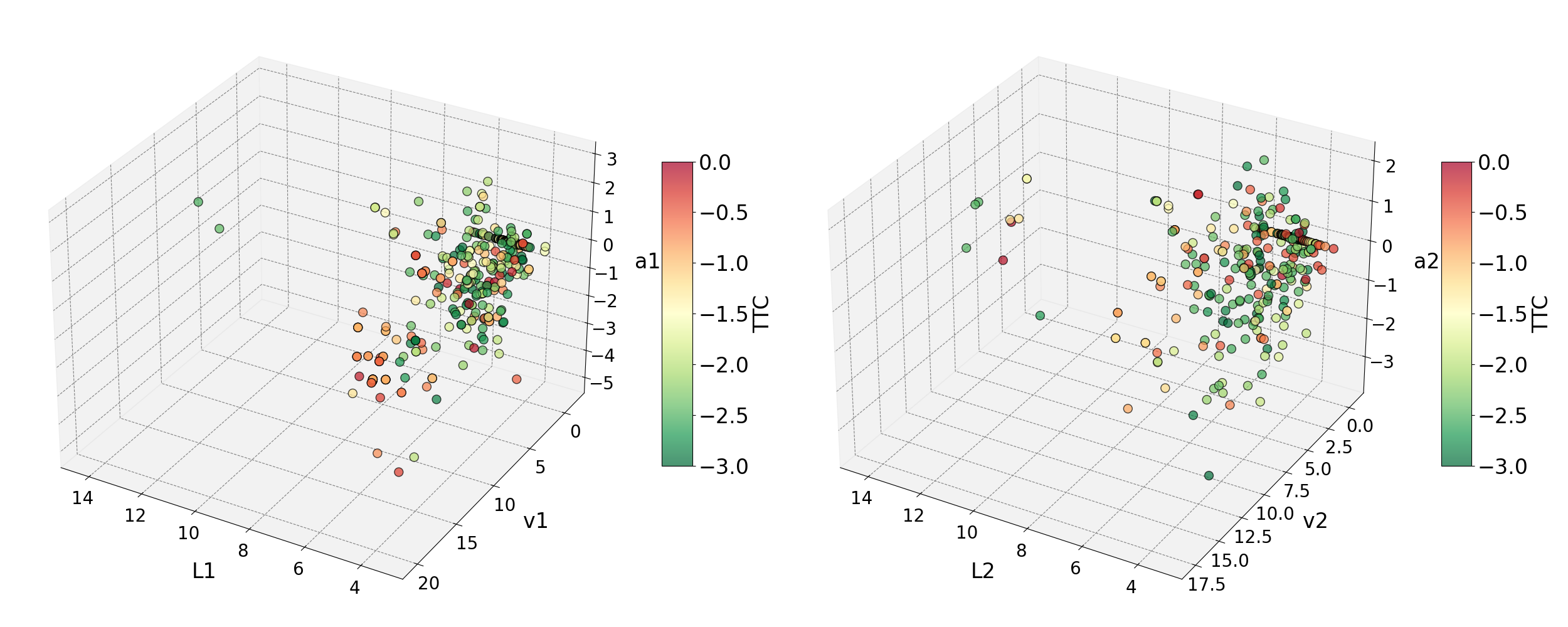}}
      \caption{Effect of exogenous variables on TTC} 
    \label{fig:8}
\end{figure}

Fig. \ref{fig:8} provides a detailed representation of the relationships between a few key parameters: traffic volume, velocity, acceleration, vehicle length —and ($TTC_{12}$). Vehicles are designated as 1 ($i$) and 2 ($j$), as referenced throughout the study, and their dynamics (For instance, v1 is $i$ vehicle's speed, and v2 is $j$ vehicle's speed). Figure \ref{fig:8}(a) shows the interaction between traffic volume, velocity, and acceleration in relation to TTC. Although no definitive trend is identified between TTC and either speed or acceleration, the findings suggest that increased traffic volume correlates with lower TTC values, implying that under congested conditions, there is reduced space and more frequent vehicle interactions. Figure \ref{fig:8}(b) illustrates the influence of vehicle length, velocity, and acceleration on TTC, indicating that shorter vehicle lengths are associated with lower TTC values, thereby enhancing the risk of collisions. This is likely because shorter vehicles are more maneuverable and capable of rapid trajectory changes, making them more prone to reduced TTC than larger vehicles. The findings emphasize the need for further investigation into the relationships between these parameters to enhance our understanding of extreme near-miss events.

\bigskip

As described in section \ref{sec3.3}, extrapolating from frequent near-miss events (2D TTC) to unobserved crash events using UGEV distribution requires an adequate sample size for model fitting. This study adopted a BM sampling approach, making it crucial to segment data into equal-sized blocks appropriately. However, there is no consensus in the literature regarding the optimal block size, with previous studies using durations ranging from 2-3 minutes to 20 minutes. Recent research by Fu et al. (\citeyear{fu2022random,fu2022bayesian}) and Zheng and Sayed (\citeyear{zhang2019real}) employed a 20-minute block size, comparing it with other sizes using local goodness-of-fit (GOF) metrics, such as the Bayesian Information Criterion (BIC). While local GOF measures provide insight into model fit, evaluating block size effects using global metrics, including mean crash estimates and their confidence intervals, is also necessary.

Identifying extreme near-miss events from AV sensor data diverges from long-duration aggregate video data methods, favoring shorter time intervals to capture detailed vehicle dynamics for the real-time framework.  In this study, each conflicting vehicle pair is treated as a block, and the minimum TTC value for each pair is used to represent extreme near-miss events, as shown in Eqn. \ref{eq:ttcmin}, previously simulator-based study by Ali et al. (\citeyear{ali2022assessing}) used each participant as a block. Selecting the minimum value for each frame could introduce bias due to the repetitive nature of conflicting pairs across frames. To avoid this, the study used each conflicting pair as a separate block, which helped minimize errors stemming from time-varying driver behavior, as suggested by Mannering  (\citeyear{mannering2018temporal}). Although some studies attribute poor GEV model performance to inadequate sample size, this issue remains contentious. While larger datasets collected over extended periods generally improve reliability, such long-term data collection can also introduce errors due to temporal variations in driver behavior.

\begin{figure}[h]
    \centering
    \setlength{\abovecaptionskip}{0pt}
    \includegraphics[width=0.7\textwidth]{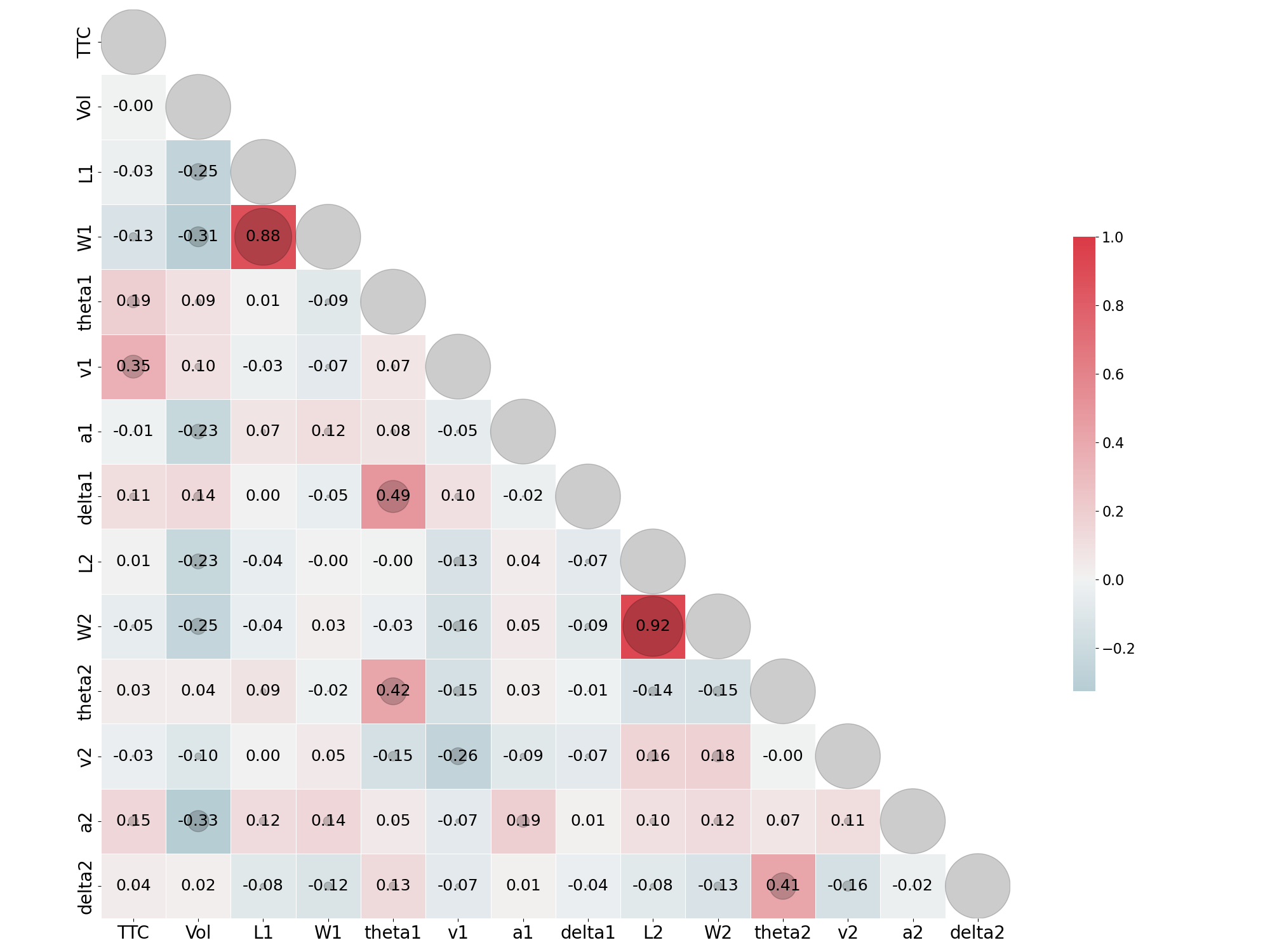}
    \caption{Correlation matrix of exogenous variables}
    \label{fig:9}
\end{figure}

As described in section \ref{sec3.3.1} the UGEV distribution accounts for unobserved heterogeneity by capturing site-specific variability through explanatory variables ($Y$), incorporated as covariates in Eqn. \ref{eq11}. A comprehensive correlation matrix was analyzed to appropriately select vehicle-specific and relevant covariates, as illustrated in ( Fig. \ref{fig:9} ). Each cell of the matrix contains correlation coefficients, ranging from -1 to 1, to indicate the strength and direction of the relationships between covariates. The analysis reveals significant correlations, such as a strong positive relationship between vehicle width and length, indicating that larger vehicles are consistently wider. Another notable positive correlation is observed between vehicle heading and steering angle. Moderate negative correlations are also found between traffic volume, vehicle speed, and acceleration, suggesting that higher traffic volumes are associated with lower speeds and reduced acceleration. These insights inform the selection of covariates in Eqn. \ref{eq11} to improve model precision.

After preparing the final dataset of extreme near-miss events, represented by negated TTC\textsubscript{min} values for each block, the data were applied to the UGEV model. Table \ref{Table:2} summarizes each site's BM frequency (extreme near-miss events) and vehicle dynamics. These are consistent with those presented in Table \ref{table:2}, supporting the robustness of the data preparation process. One of the primary objectives of this study was to incorporate individual vehicle dynamics into the HBSRP model, thereby developing a non-stationary model capable of capturing site-specific variability. The influence of speed, volume, acceleration, and deceleration on crash risk has been well-established in the literature. Increased traffic volume is known to elevate crash risk, and speed has been found to correlate strongly with heightened near-miss risk (\citep{abdel2008assessing}). Controlling input covariates of vehicle dynamics, both acceleration and deceleration, also correlate with increasing near-miss crash risk, and previous studies investigated (\citep{desai2021correlating,jun2007relationships, tak2015development}). Similarly, higher steering angles, indicative of swerving, are associated with increased near-miss risk due to sudden lateral movements that may catch following vehicles off guard, often resulting in secondary collisions (\citep{gilbert2021multi}). Thus, these vehicle dynamics were incorporated as covariates within the HBSRP model to achieve precise real-time near-miss crash risk estimation. The analysis across the six sites revealed block maxima frequencies ranging from 45 to 108, thereby meeting the minimum threshold of 30 observations for robust block maxima based sampling, as recommended by Zheng et al. (\citeyear{zheng2014freeway}). Specific insights from individual sites indicate varying traffic dynamics and potential risk levels. For instance, Site 5 shows Vehicle-1 with a high speed of 7.78 m/s and a deceleration rate of -0.56 m/s², suggesting this location is prone to frequent interactions and abrupt maneuvers, thereby indicating a potentially high-risk environment. In contrast, Site 3 demonstrates Vehicle-2 with a high speed of 9.7 m/s but minimal deceleration of -0.17 m/s², which means less frequent but more severe interactions. Sites 1 and 2 exhibit balanced vehicle dynamics characterized by moderate speeds, mixed acceleration, and deceleration, reflecting typical urban traffic conditions with diverse movement behaviors. Site 4 also demonstrates balanced dynamics, with similar acceleration patterns for both vehicles, suggesting synchronized traffic flow. Conversely, Site 6, characterized by lower vehicle speeds, indicates a stable yet less dynamic traffic environment. These findings highlight the importance of incorporating detailed vehicle dynamics into the modeling framework to reflect site-specific crash risks accurately.

\begin{table}[h]
\centering
\caption{Summary statistics of each site}
\label{Table:2}
\begin{tabular}{l l l l  l l l l l l}
\hline
Site & Block maxima & \multicolumn{3}{c}{TTC\textsubscript{min} (s)} & \multicolumn{2}{c}{Vehicle-1} & \multicolumn{2}{c}{Vehicle-2} \\
     & (Frequency) & avg & min & max &   spd (m/s) & acc/dec (m/s\(^2\)) & spd (m/s) & acc/dec (m/s\(^2\)) \\ \hline
1    & 58 & 1.96 & 0.1 & 3.00 & 3.89 & -0.255 & 2.61 & -0.15 \\
2    & 95 & 2.18 & 0.1 & 3.00 & 2.4 & 0.06 & 3.33 & -0.09 \\
3    & 45 & 1.21 & 0.15 & 2.846 & 9.7 & -0.17 & 2.45 & 0.07 \\
4    & 61 & 1.36 & 0.1 & 2.92 & 4.9 & 0.57 & 4.55 & 0.66 \\
5    & 108 & 1.44 & 0.1 & 2.97& 7.78 & -0.56 & 2.24 & -0.48 \\
6    & 55 & 1.58 & 0.3 & 2.53& 1.13 & -0.03 & 0.87 & 0.04 \\
\hline
\end{tabular}
\end{table}

\subsection{Model estimation inference \label{sec4.2}}

Several UGEV models were developed for six sites across three cities to estimate crash risks based on extreme 2D TTC near-miss events. Rather than modeling each site individually, the UGEV distribution was fitted collectively to data from grouped sites within each city, effectively treating them as a network. This network-based approach addressed the issue of sample scarcity by pooling data at the city level, thus enhancing the robustness of the analysis. The modeling relied on the OpenBUGS software, efficiently handling the complex computations needed through Markov Chain Monte Carlo (MCMC) algorithms. Given the potential biases and auto-correlation in MCMC sampling, extensive iterations were performed to mitigate these issues and ensure reliable parameter estimates. Multiple simulation chains were employed, each starting with different initial values to enhance convergence reliability. Specifically, two independent chains were executed for 50,000 iterations for each parameter, with the first 20,000 iterations discarded as burn-in, leaving the final 30,000 iterations to be used for posterior estimation.

Convergence of the models was verified through both visual and quantitative diagnostics. Trace plots were visually inspected to ensure consistent progression of the chains from varied starting points, and the Brooks-Gelman-Rubin (BGR) statistic was calculated for each parameter. (\citep{el2009urban}). The BGR statistic stabilized around 1, specifically below the threshold of 1.1, indicating effective convergence (\citep{gelman1992inference}).

Following the model development, the DIC values were used to evaluate four distinct UGEV models, as summarized in Tables \ref{Table:3}, \ref{Table:4}, and \ref{Table:5}. These models included both stationary and non-stationary approaches. The stationary models comprised fixed parameter models, assuming uniformity across all observations, and random parameter models, which allowed variability across different sites. In contrast, the non-stationary models accounted for dynamic behavior by incorporating variability using covariates in the location and scale parameters. Specifically, one non-stationary model parameterized both the location and scale parameters, while another extended this framework by including random site-based effects in the location parameter and a scale-parameterized component. The evaluation indicated that non-stationary models, particularly those allowing parameterization of both location and scale, outperformed stationary models significantly regarding model fit. This enhanced performance is attributed to the strategic integration of covariates, which effectively captured site-specific variations and the intricate interactions in the dataset. Such modeling provided more profound insights into traffic interactions, improving the GOF. Among the non-stationary models, the best-performing approach, evidenced by the lowest DIC value, included random parameters across different sites and covariate integration in the location and scale parameters. This model demonstrated the highest robustness and accuracy in crash risk estimation and was thus selected for further analysis.

Several exogenous covariates, such as volume, object length, width, speed, acceleration, heading, and steering angle, were initially included in the modeling process. However, after assessing their statistical significance, it was determined that some covariates, namely, object dimensions, heading, and steering angle, did not contribute meaningfully and were therefore excluded from the final model. The volume covariate was also removed due to its high correlation with vehicle speed and acceleration, which could result in multicollinearity issues. This refinement resulted in a more parsimonious and robust model, enhancing both the reliability and accuracy of crash risk estimates. By focusing only on the most impactful variables, the model's ability to capture the nuanced dynamics of crash risk at each site was significantly improved.

\begin{table}[h]
\centering
\caption{Model estimation results for sites 1\&2 (San Francisco)}
\label{Table:3}
\resizebox{\textwidth}{!}{
\begin{tabular}{l l l l l l l l l l l l l l l l l l}
\hline
\multicolumn{2}{c}{\text{Model parameters}} & \multicolumn{8}{c}{\text{Stationary}} & \multicolumn{8}{c}{\text{Non-Stationary}} \\
\cline{3-18}
\hline
 & &  \multicolumn{4}{c}{\text{FP}} & \multicolumn{4}{c}{\text{RP}} & \multicolumn{4}{c}{\text{FP}} & \multicolumn{4}{c}{\text{RP}} \\ 
 & &  \text{Mean} & \text{SD} & \text{2.5\%} & \text{97.5\%} & \text{Mean} & \text{SD} & \text{2.5\%} & \text{97.5\%} & \text{Mean} & \text{SD} & \text{2.5\%} & \text{97.5\%} & \text{Mean} & \text{SD} & \text{2.5\%} & \text{97.5\%} \\
\hline
$\mu$ & $\alpha_{\mu_0}$ & -2.499 & 0.04 & -2.576 & -2.418 &  &  &  &  & -2.489& 0.037 & -2.56 & -2.413 & & & &\\
 & $\alpha_{\mu_{0[1]}}$ &  &  & & & -2.485 & 0.103 & -2.667 & -2.268 & & & & & -2.605 & 0.064 & -2.718 & -2.467\\
 & $\alpha_{\mu_{0[2]}}$ &  &  & & & -2.491 & 0.039 & -2.565 & -2.411 & & & & & -2.472 & 0.035 & -2.538 & -2.402  \\
 & $\alpha_{\mu_{\text{spd\_veh1}}}$ &  &  &  &  &  &  &  &  &  &  &  &  &  & \\
 & $\alpha_{\mu_{\text{spd\_veh2}}}$  &  &  &  & &  &  &  & & -0.047 & 0.008 & -0.0613 & -0.0281& -0.028 & 0.008 & -0.044 & -0.013\\
 & $\alpha_{\mu_{\text{acc\_veh1}}}$  &  &  &  &  &  &  &  &  &  &  &  &  &  -0.049& 0.024 &  -0.094&-0.030 \\
 & $\alpha_{\mu_{\text{acc\_veh2}}}$  &  &  &  &  & &  &  &  &  &  &  &  &  &  &  & \\
 
$\vartheta = \log(\sigma)$ & $\alpha_{\vartheta_0}$ & -0.846 & 0.08 & -1.0 & -0.688 &  &  & & &-0.888&0.0787&-1.04&-0.732& & & & \\
 & $\alpha_{\vartheta_{0[1]}}$ &  &  & & & -0.566 & 0.166 & -0.901 & --0.254 &  & & & & -0.795 & 0.158 & -1.098 & -0.478\\
 & $\alpha_{\vartheta_{0[2]}}$ &  &  & & & -1.077 & 0.095 & -1.259 & -0.885 &  & & & & -1.127 & 0.091 & -1.3 & -0.941 \\
 & $\alpha_{\vartheta_{\text{spd\_veh1}}}$ &  &  &  &  &  &  &  &  &  &  &  &  &  &  &  & \\
 & $\alpha_{\vartheta_{\text{spd\_veh2}}}$ &  &  &  &  &  &  &  &  & -0.074 & 0.027 & -0.124 & -0.019 & -0.104 & 0.026 & -0.152 & -0.053\\
 & $\alpha_{\vartheta_{\text{acc\_veh1}}}$ &  &  &  &  &  &  &  &  &  &  &  &  &  &  &  & \\
 & $\alpha_{\vartheta_{\text{acc\_veh2}}}$ &  &  &  &  &  &  &  &  & 0.275 & 0.113 & 0.002 & 0.059 & 0.298 & 0.102 & 0.105 & 0.507\\

$\zeta$ & $\alpha_{\xi_0}$ & 0.299 & 0.079 & -0.151 & -0.4634 & & &&& 0.286 & 0.074 & 0.149 & 0.438 & & & &\\
 & $\alpha_{\xi_{0[1]}}$ &  &  & & & 0.329& 0.226 & 0.031 & 0.813 &  &  & & & 0.559 & 0.183 & 0.203 & 0.916 \\
 & $\alpha_{\xi_{0[2]}}$ &  &  & & & 0.275 & 0.074 & 0.139 & 0.429 &  & & & & 0.220 & 0.069 & 0.094 & 0.367 \\
$\bar{D}$ & &  & & 274.5 & & & &260.8 & & & & 261.6 & & & \textbf{241.3} & \\
$p_D$ & &  & & 1.54 & & & & 2.953 & & & & 3.176 & & & 5.645 & \\
DIC & & & & 276.1 & & & & 263.7 & & & & 264.8 & & & \textbf{246.95} & \\
\hline
\end{tabular}
}
\end{table}

\begin{table}[h]
\centering
\caption{Model estimation results for sites 3\&4 (Phoenix)}
\label{Table:4}
\resizebox{\textwidth}{!}{
\begin{tabular}{l l l l l l l l l l l l l l l l l l}
\hline
\multicolumn{2}{c}{\text{Model parameters}} & \multicolumn{8}{c}{\text{Stationary}} & \multicolumn{8}{c}{\text{Non-Stationary}} \\
\cline{3-18}
\hline
 & &  \multicolumn{4}{c}{\text{FP}} & \multicolumn{4}{c}{\text{RP}} & \multicolumn{4}{c}{\text{FP}} & \multicolumn{4}{c}{\text{RP}} \\ 
 & &  \text{Mean} & \text{SD} & \text{2.5} & \text{97.5} & \text{Mean} & \text{SD} & \text{2.5} & \text{97.5} & \text{Mean} & \text{SD} & \text{2.5} & \text{97.5} & \text{Mean} & \text{SD} & \text{2.5} & \text{97.5} \\
\hline
$\mu$ & $\alpha_{\mu_0}$ & -1.466 & 0.086 & -1.641 & -1.302 &  &  &  &  & -1.535 & 0.057 & -1.647 & -1.424 & & & &\\
 & $\alpha_{\mu_{0[1]}}$ &  &  & & & -1.363 & 0.124 & -1.616 & -1.132 & & & & & -1.349 & 0.072 & -1.497 & -1.212\\
 & $\alpha_{\mu_{0[2]}}$ &  &  & & & -1.56 & 0.122 & -1.807 & -1.327 & & & & & -1.677 & 0.084 & -1.843 & -1.512  \\
 & $\alpha_{\mu_{\text{spd\_veh1}}}$ &  &  &  &  &  &  &  &  & 0.101  & 0.014 & 0.073  & 0.128 &0.093  &0.012& 0.071 & 0.116 \\
 & $\alpha_{\mu_{\text{spd\_veh2}}}$  &  &  &  & &  &  &  & & 0.1 & 0.013 & 0.073 & 0.125& 0.083 & 0.012 &0.061  & 0.107\\
 & $\alpha_{\mu_{\text{acc\_veh1}}}$  &  &  &  &  &  &  &  &  &  &  &  &  &  & & & \\
 & $\alpha_{\mu_{\text{acc\_veh2}}}$  &  &  &  &  & &  &  &  &  &  &  &  &  &  &  & \\

$\vartheta = \log(\sigma)$ & $\alpha_{\vartheta_0}$ & -0.207 & 0.084 & -0.359 & -0.031 &  &  & & &-0.519& 0.068 & -0.651 & -0.384 & & & & \\
 & $\alpha_{\vartheta_{0[1]}}$ &  &  & & & -0.285 & 0.127 & -0.507 & -0.014 &  & & & & -0.755 &0.136 & -1.007  & -0.476 \\
 & $\alpha_{\vartheta_{0[2]}}$ &  &  & & & -0.152 & 0.111 & -0.352 & 0.088 &  & & & & -0.418  &0.107 & -0.620  &-0.198  \\
 & $\alpha_{\vartheta_{\text{spd\_veh1}}}$ &  &  &  &  &  &  &  &  &  &  &  &  &  &  &  & \\
 & $\alpha_{\vartheta_{\text{spd\_veh2}}}$ &  &  &  &  &  &  &  &  &  & &  &  & -0.055 & 0.026 & -0.104 & -0.003\\
 & $\alpha_{\vartheta_{\text{acc\_veh1}}}$ &  &  &  &  &  &  &  & & -0.359 & 0.068 & -0.493 &-0.227 &-0.393 & 0.073 & -0.538 & -0.249\\
 & $\alpha_{\vartheta_{\text{acc\_veh2}}}$ &  &  &  &  &  &  &  &  & 0.146 & 0.054 & 0.036 & 0.251 & 0.293  & 0.081 & 0.135 &0.452 \\
 
$\zeta$ & $\alpha_{\xi_0}$ & -0.546 & 0.067 & -0.684 & -0.415 & & &&& 0.015 & 0.015 & 0.0003 & 0.055 & & & &\\
 & $\alpha_{\xi_{0[1]}}$ &  &  & & & -0.573 & 0.094 & -0.765 & -0.393 &  &  & & & 0.043 & 0.040 & 0.001 & 0.149 \\
 & $\alpha_{\xi_{0[2]}}$ &  &  & & & -0.506 & 0.102 & -0.706 & -0.300 &  & & & & 0.027 & 0.027 & 0.0006 & 0.101 \\
$\bar{D}$ & &  & & 223.5 & & & & 224.0 & & & & 202.4 & & & \textbf{195.551} & \\
$p_D$ & &  & & 1.623 & & & & 3.396 & & & & 4.735 & & & 5.849 & \\
DIC & & & & 225.1 & & & & 227.4 & & & & 207.2 & & & \textbf{201.4} & \\
\hline
\end{tabular}
}
\end{table}

\begin{table}[h]
\centering
\caption{Model estimation results for sites 5\&6 (Los Angeles)}
\label{Table:5}
\resizebox{\textwidth}{!}{
\begin{tabular}{l l l l l l l l l l l l l l l l l l}
\hline
\multicolumn{2}{c}{\text{Model parameters}} & \multicolumn{8}{c}{\text{Stationary}} & \multicolumn{8}{c}{\text{Non-Stationary}} \\
\cline{3-18}
\hline
 & &  \multicolumn{4}{c}{\text{FP}} & \multicolumn{4}{c}{\text{RP}} & \multicolumn{4}{c}{\text{FP}} & \multicolumn{4}{c}{\text{RP}} \\ 
 & &  \text{Mean} & \text{SD} & \text{2.5} & \text{97.5} & \text{Mean} & \text{SD} & \text{2.5} & \text{97.5} & \text{Mean} & \text{SD} & \text{2.5} & \text{97.5} & \text{Mean} & \text{SD} & \text{2.5} & \text{97.5} \\
\hline
$\mu$ & $\alpha_{\mu_0}$ & -1.885 & 0.062 & -2.005 & -1.764 &  &  &  &  & -1.854 & 0.059 & -1.969 & -1.737 & & & &\\
 & $\alpha_{\mu_{0[1]}}$ &  &  & & & -1.889 & 0.085 & -2.054 & -1.72 & & & & & -1.866 & 0.084 & -2.029 & -1.698\\
 & $\alpha_{\mu_{0[2]}}$ &  &  & & & -1.875 & 0.077 & -2.025 & -1.722 & & & & & -1.857 & 0.074 & -2.001 &-1.709   \\
 & $\alpha_{\mu_{\text{spd\_veh1}}}$ &  &  &  &  &  &  &  &  & 0.046  & 0.011 & 0.025  & 0.067 &0.048 &0.012 & 0.024 &0.072 \\
 & $\alpha_{\mu_{\text{spd\_veh2}}}$  &  &  &  & &  &  &  & & & &  & &  &  &  & \\
 & $\alpha_{\mu_{\text{acc\_veh1}}}$  &  &  &  &  &  &  &  &  &  &  &  &  &  & & & \\
 & $\alpha_{\mu_{\text{acc\_veh2}}}$  &  &  &  &  & &  &  &  & 0.181  &  0.072&0.044  & 0.328 & 0.018 & 0.020 & 0.008 &0.057 \\

$\vartheta = \log(\sigma)$ & $\alpha_{\vartheta_0}$ & -0.309 & 0.061 & -0.427 & -0.188 &  &  & & &-0.344& 0.059 & -0.456 & -0.227 & & & & \\
 & $\alpha_{\vartheta_{0[1]}}$ &  &  & & & -0.19 & 0.0812 &-0.249  & -0.049 &  & & & & -0.209 &0.075 &-0.349   &-0.058  \\
 & $\alpha_{\vartheta_{0[2]}}$ &  &  & & &  -0.65 &0.11 & -0.910& -0.540&  &	  & & & -0.689  & 0.107& -0.890  &-0.470   \\
 & $\alpha_{\vartheta_{\text{spd\_veh1}}}$ &  &  &  &  &  &  &  &  &  &  &  &  &  &  &  & \\
 & $\alpha_{\vartheta_{\text{spd\_veh2}}}$ &  &  &  &  &  &  &  &  &  & &  &  &  &  &  & \\
 & $\alpha_{\vartheta_{\text{acc\_veh1}}}$ &  &  &  &  &  &  &  & &  &  &  & & &  &  & \\
 & $\alpha_{\vartheta_{\text{acc\_veh2}}}$ &  &  &  &  &  &  &  &  &  &  &  &  &  & & & \\

$\zeta$ & $\alpha_{\xi_0}$ & 0.02 & 0.019 & 0.0005 & 0.072 & & &&& 0.015 & 0.015 & 0.0003 & 0.053 & & & &\\
 & $\alpha_{\xi_{0[1]}}$ &  &  & & & 0.031 & 0.031 & 0.0007 & 0.112 &  &  & & & 0.024 & 0.024 & 0.0061 & 0.089 \\
 & $\alpha_{\xi_{0[2]}}$ &  &  & & & 0.061 & 0.059 & 0.002 & 0.219 &  & & & & 0.055 & 0.053 & 	0.002 & 0.197 \\
$\bar{D}$ & &  & & 402.1 & & & & 392.7 & & & & 383.3 & & & \textbf{374.95} & \\
$p_D$ & &  & & 1.966 & & & & 3.929 & & & & 2.979 & & & 4.749 & \\
DIC & & & & 404.1 & & & & 396.6 & & & & 386.2 & & & \textbf{379.7} & \\
\hline
\end{tabular}
}
\end{table}

Incorporating the selected covariates, the non-stationary model with random parameters (HBSRP) for both location and scale provided enhanced flexibility by addressing both temporal variability and random effects. The comparative analysis shown in Tables \ref{Table:3}, \ref{Table:4}, and \ref{Table:5}) highlights the superiority of non-stationary models with random parameters in accurately capturing the complexities of the dataset, resulting in the most reliable crash risk estimates. The assessment of model fit confirmed that the HBSRP model achieved the lowest DIC values, indicating its superior performance compared to other models. Specifically, the DIC values for the best-fitted models were 246.95 for San Francisco, 201.4 for Phoenix, and 379.7 for Los Angeles, with a 6.43-10.56\% decrease in DIC, demonstrating the robustness of the HBSRP approach in capturing site-specific crash risk dynamics.

\begin{figure}[h!]
    \centering
    \setlength{\abovecaptionskip}{0pt}
    \subcaptionbox{FP with stationary and non-stationary}
    {\includegraphics[width=0.45\textwidth]{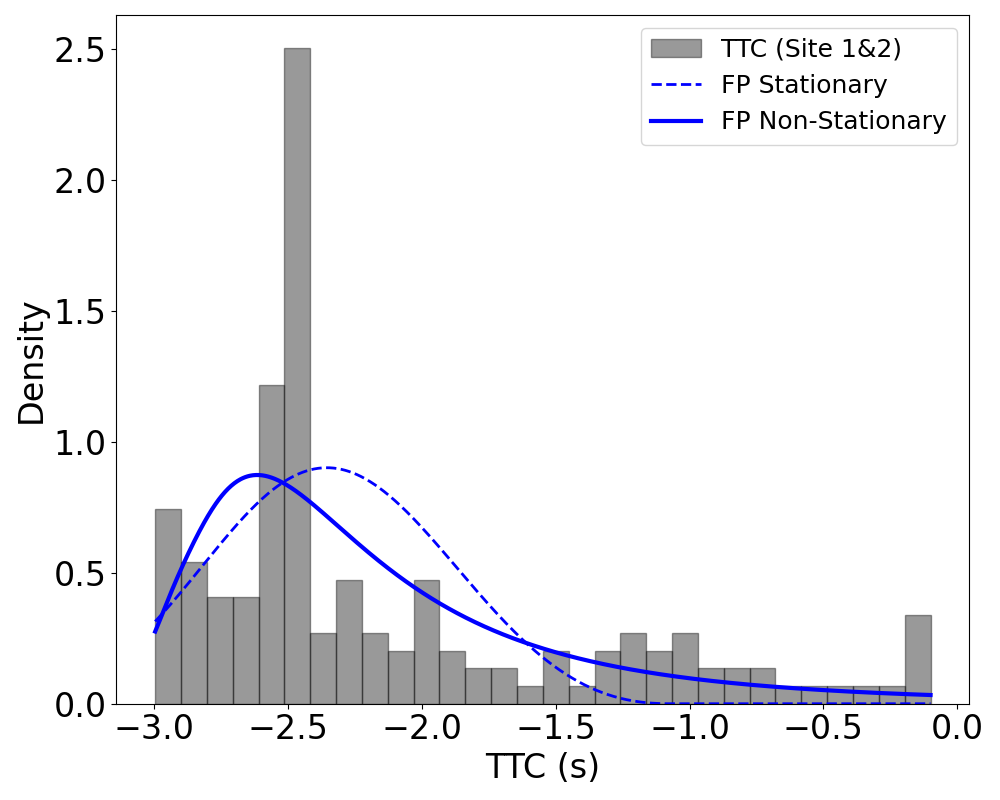}}
    \subcaptionbox{RP with stationary and non-stationary}{\includegraphics[width=0.45\textwidth]{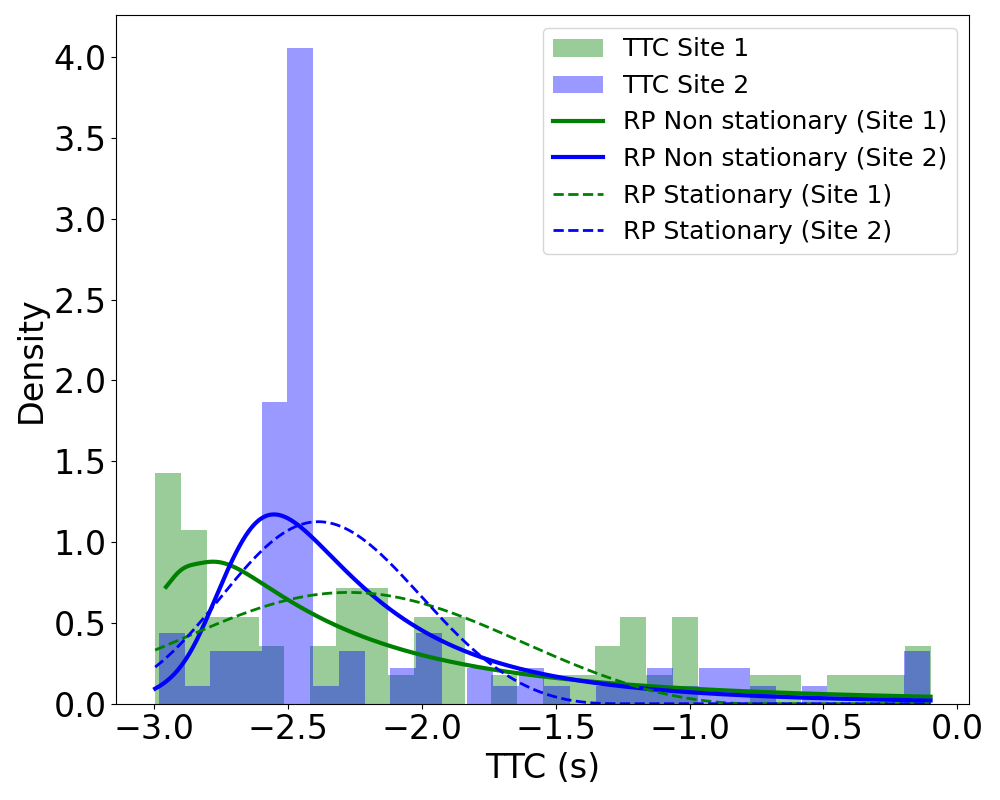}}
    \caption{Posterior predictive model check for San Francisco sites}
    \label{fig:10}
\end{figure}

\begin{figure}[h!]
    \centering
    \setlength{\abovecaptionskip}{0pt}
    \subcaptionbox{FP with stationary and non-stationary}
    {\includegraphics[width=0.45\textwidth]{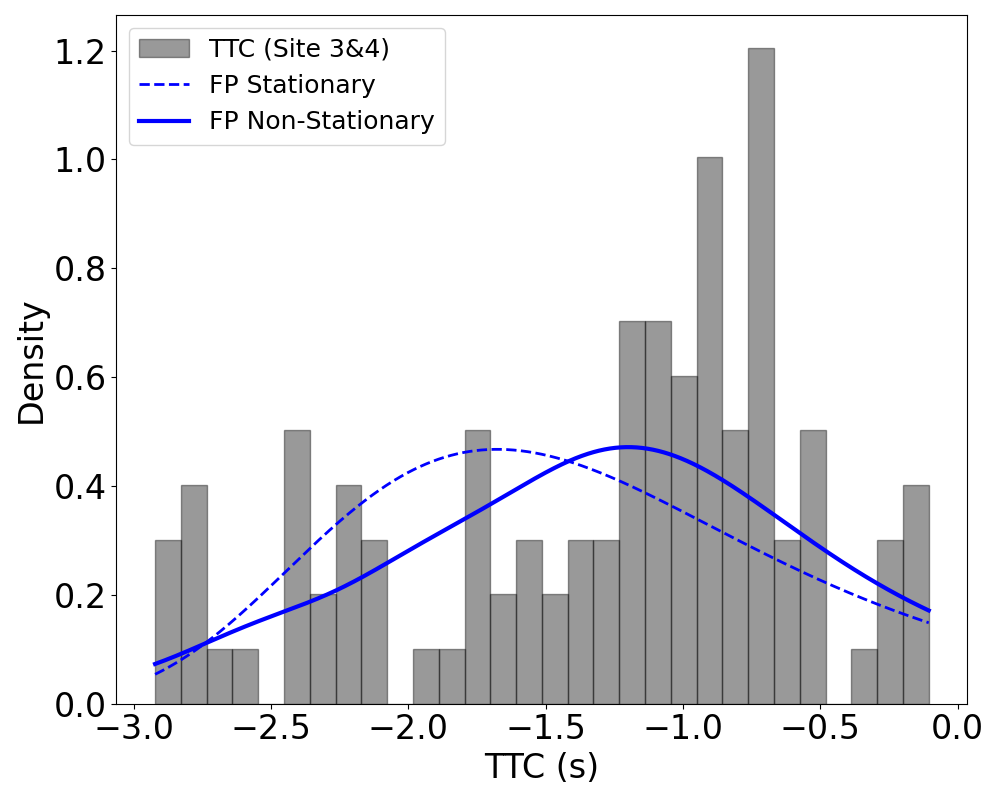}}
    \subcaptionbox{RP with stationary and non-stationary}{\includegraphics[width=0.45\textwidth]{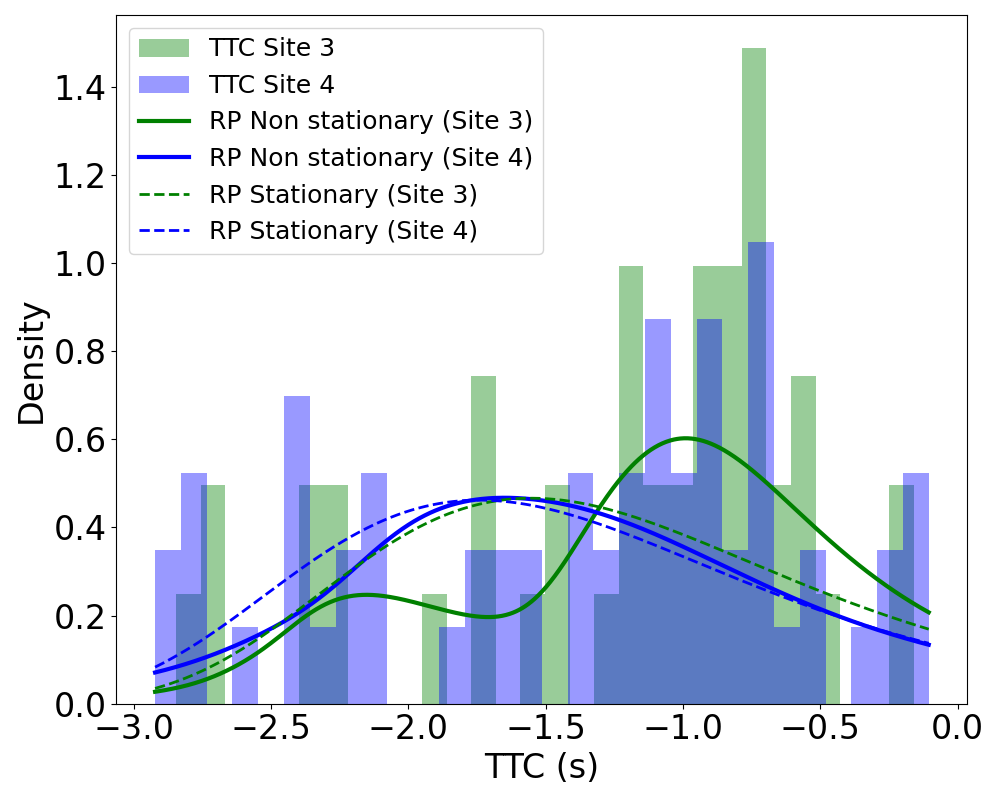}}
    \caption{Posterior predictive model check for Phoenix sites}
    \label{fig:11}
\end{figure}

\begin{figure}[h!]
    \centering
    \setlength{\abovecaptionskip}{0pt}
    \subcaptionbox{FP with stationary and non-stationary}
    {\includegraphics[width=0.45\textwidth]{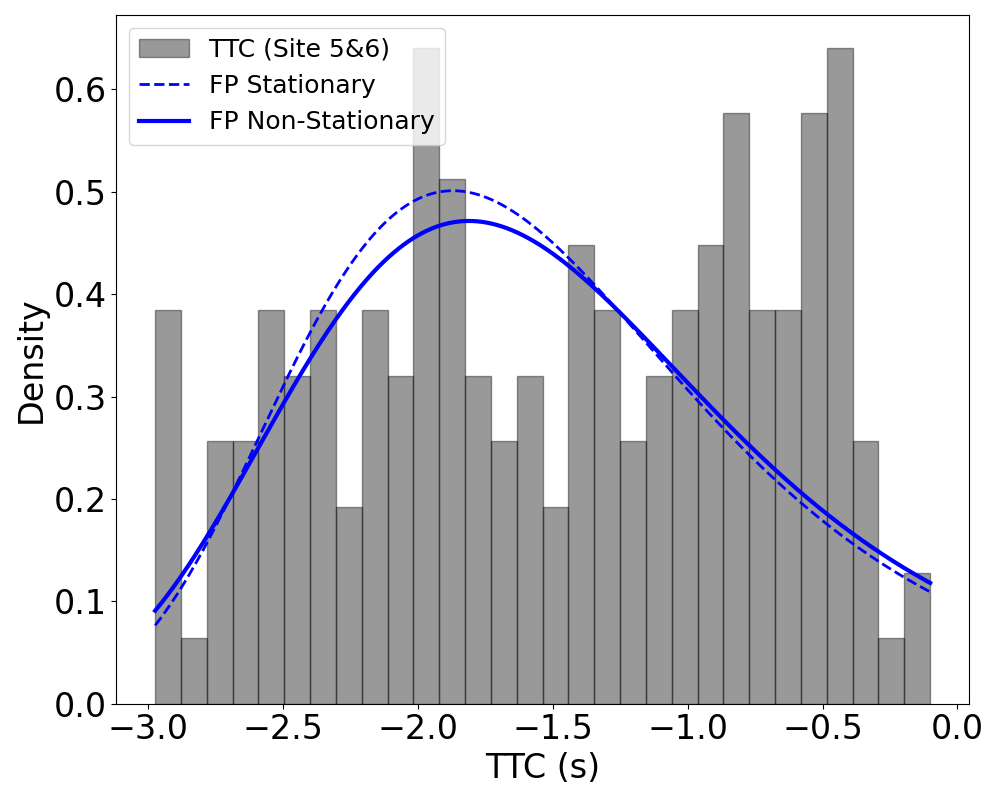}}
    \subcaptionbox{RP with stationary and non-stationary}{\includegraphics[width=0.45\textwidth]{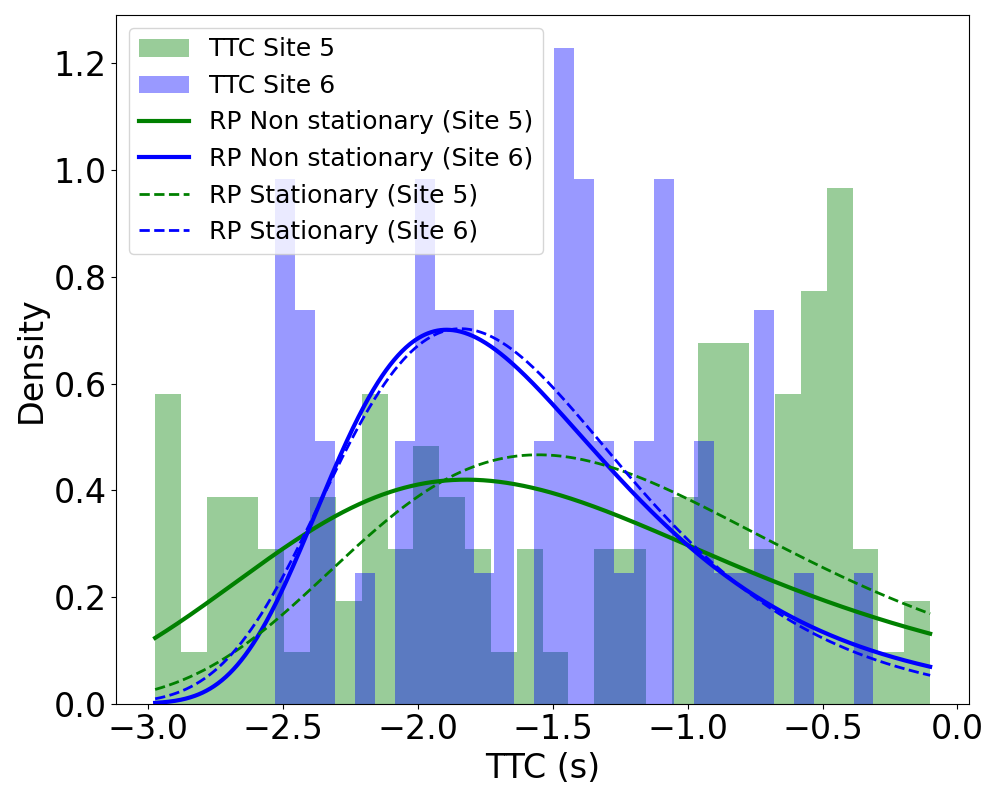}}
    \caption{Posterior predictive model check for Los Angeles sites}
    \label{fig:12}
\end{figure}

Once proper sampling was confirmed with minimal autocorrelation; the model fit was evaluated using posterior predictive checks by comparing model-generated data with observed values (\citep{gelman1996posterior}). This assessment involved generating data from samples drawn from the posterior distribution using the likelihood function, allowing for a robust evaluation of the model's predictive accuracy. Figures \ref{fig:10}, \ref{fig:11}, and \ref{fig:12} illustrate the results of these checks, revealing that non-stationary models consistently outperformed stationary models. This superior performance underscores the importance of accounting for random effects and temporal variability in crash risk estimation. Notably, the HBSRP model demonstrated the closest alignment with observed data, especially in predicting extreme values, indicating a superior fit. Furthermore, covariates signs and magnitudes of mean estimates presented in Tables \ref{Table:3}, \ref{Table:4}, and \ref{Table:5} offer their influence on crash risk.

\bigskip
The HBSRP model exhibited superior performance for the San Francisco dataset compared to alternative models, as evidenced by a DIC difference greater than ten. Similar findings were observed for the Phoenix and Los Angeles datasets, where the differences in DIC between fixed parameter (FP) models and random parameter (RP) non-stationary models exceeded five. Despite these differences, the RP non-stationary model (HBSRP) consistently yielded the lowest DIC values across all cities, underscoring its superior fit. This improved performance can be attributed to the model's capability to effectively capture unobserved heterogeneity by allowing parameters to vary across observations, accommodating the inherent variability in vehicle dynamics. The non-stationary framework of the HBSRP model facilitates the adaptation to both temporal and spatial fluctuations, thereby offering a more realistic representation of complex, real-world traffic conditions. Furthermore, the application of MCMC algorithms enhances the precision of parameter estimation, mainly when vehicle-specific dynamics are included as covariates. This inclusion significantly improves the reliability and accuracy of crash risk predictions. The hierarchical Bayesian structure of the model further refines its ability to distinguish between within-site and between-site variability, thereby capturing the nuances of site-specific effects. Additionally, using extreme value distributions enables the model to focus on rare but critical traffic events, essential for evaluating high-risk scenarios and ensuring robust crash risk assessments. These attributes of the HBSRP model underscore its suitability for modeling complex traffic environments and generating precise risk estimates that are instrumental for targeted safety interventions.

The analysis of model estimation results indicates significant associations between vehicle-specific dynamics and the UGEV distribution parameters for 2D TTC across different city datasets. In the San Francisco dataset (Table:\ref{Table:3}), vehicle-2's speed and vehicle-1's acceleration are significantly associated with the location parameter, while both the speed and acceleration vehicle-2 exhibit a significant relationship with the scale parameter. Vehicle-1 and vehicle-2 are illustrated in Fig. \ref{fig:3} for reference. In Phoenix (Table: \ref{Table:4}), vehicle-1 and vehicle-2 speeds significantly impact the location parameter. In addition, the speed and acceleration of vehicle-1, along with the speed of vehicle-2, are found to be influential in determining the scale parameter. In the case of Los Angeles (Table: \ref{Table:5}), vehicle-1's speed and vehicle-2's acceleration are linked to the location parameter, while no significant associations were observed for the scale parameter. The HBSRP model results further validate the appropriateness of the variance estimates for the intercept random coefficient within the GEV distribution for TTC. This applies to the location, shape, and scale parameters, confirming their statistical reliability.

The comprehensive analysis yields both theoretical insights and practical implications. Notably, non-stationary models with random parameters outperform their stationary counterparts, underscoring the necessity of incorporating temporal variations and random effects when predicting near-miss and crash risks. The unique near-miss patterns observed across different sites emphasize the need for site-specific safety interventions. The GEV model, integrated with a hierarchical Bayesian structure, effectively captures the extreme values or tail behavior of extreme near-miss distributions critical for assessing each scenario. Significant distinctions between stationary and non-stationary models demonstrate the importance of accounting for time-varying factors. This study considers vehicle-specific dynamics in the short term in traffic safety assessments. These models achieve greater flexibility and enhanced performance by incorporating random parameters, effectively capturing the unobserved heterogeneity across sites. Together, these findings establish a robust framework for comprehending real-time near-miss crash risk and guiding the implementation of proactive safety measures.

\subsection{Model validation}\label{sec4.3}

Since the study dataset is anonymous, it lacks specific geolocation, date, and time information. Consequently, no definitive ground truth is available for identifying real extreme events or actual crashes in the study area. Despite this limitation, evaluating the HBSRP model's performance is crucial. A cross-fold validation approach was adopted to address this. We split the dataset into training and test datasets for model validation. The training dataset is used to calibrate the parameters, while the test dataset is used to assess the model's performance. This estimation allows us to compute the expected number of extreme near misses for different severity levels and compare these estimates with the observed near miss frequencies to assess the model’s predictive accuracy.

In the validation process, the study estimates the frequency of near misses (C) at various severity thresholds (\(\lambda\)) using Eqn. \ref{eq22}, and derived from Eqn. \ref{eq15} . The estimation is obtained by summing the probabilities of block maxima (\(X_n\)) exceeding the threshold (\(\lambda\)) for each block. The estimated frequencies are calculated by:

\begin{equation} \label{eq22}
C_{m\lambda} = \begin{cases} 
\sum_{n=1}^{k} \left(1 - \exp\left[-\left(1 - \xi_n \frac{\lambda - \mu_n}{\exp(\vartheta_n)}\right)^{-1/\xi_n}\right]\right), & \text{for } \xi_n \neq 0 \\
\sum_{n=1}^{k} \left(1 - \exp\left[-\exp\left(\frac{\lambda - \mu_n}{\exp(\vartheta_n)}\right)\right]\right), & \text{for } \xi_n = 0
\end{cases}
\end{equation}

\begin{figure}[h!]
    \centering
    \setlength{\abovecaptionskip}{0pt}
    \subcaptionbox{Site-1}
    {\includegraphics[width=0.4\textwidth]{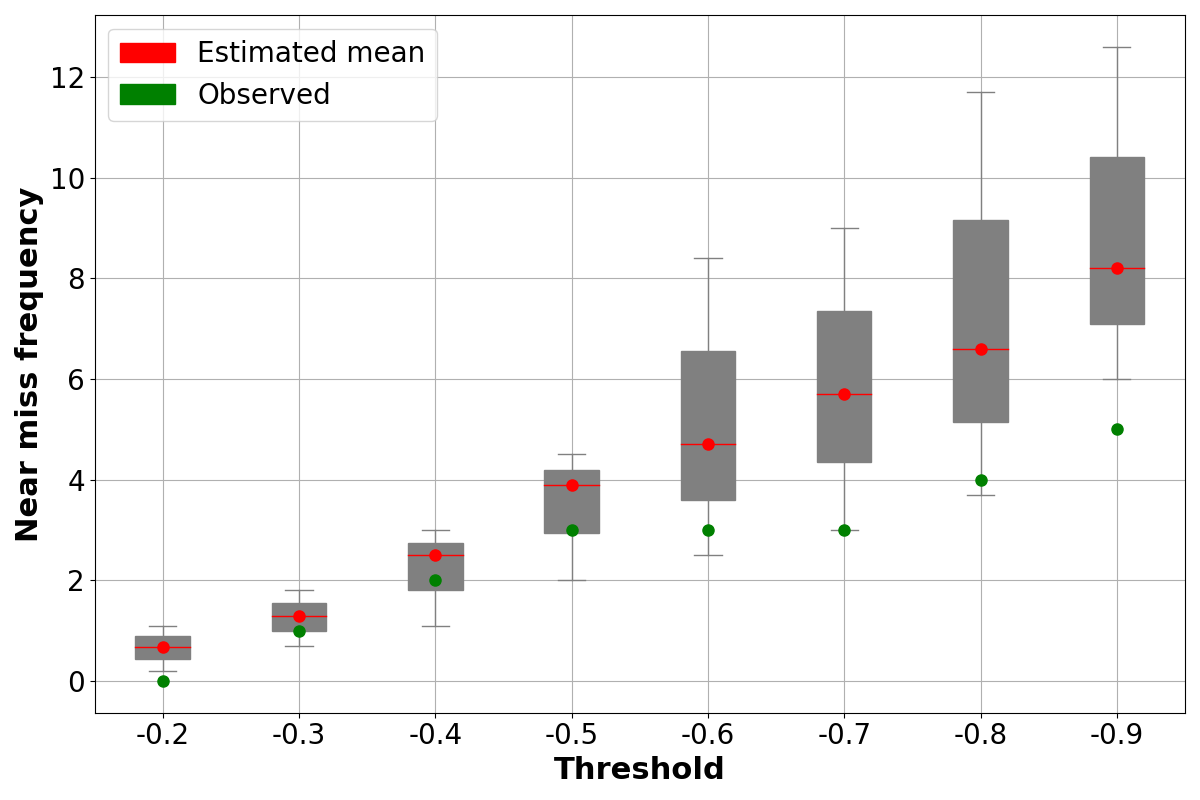}}
    \subcaptionbox{Site-2}{\includegraphics[width=0.4\textwidth]{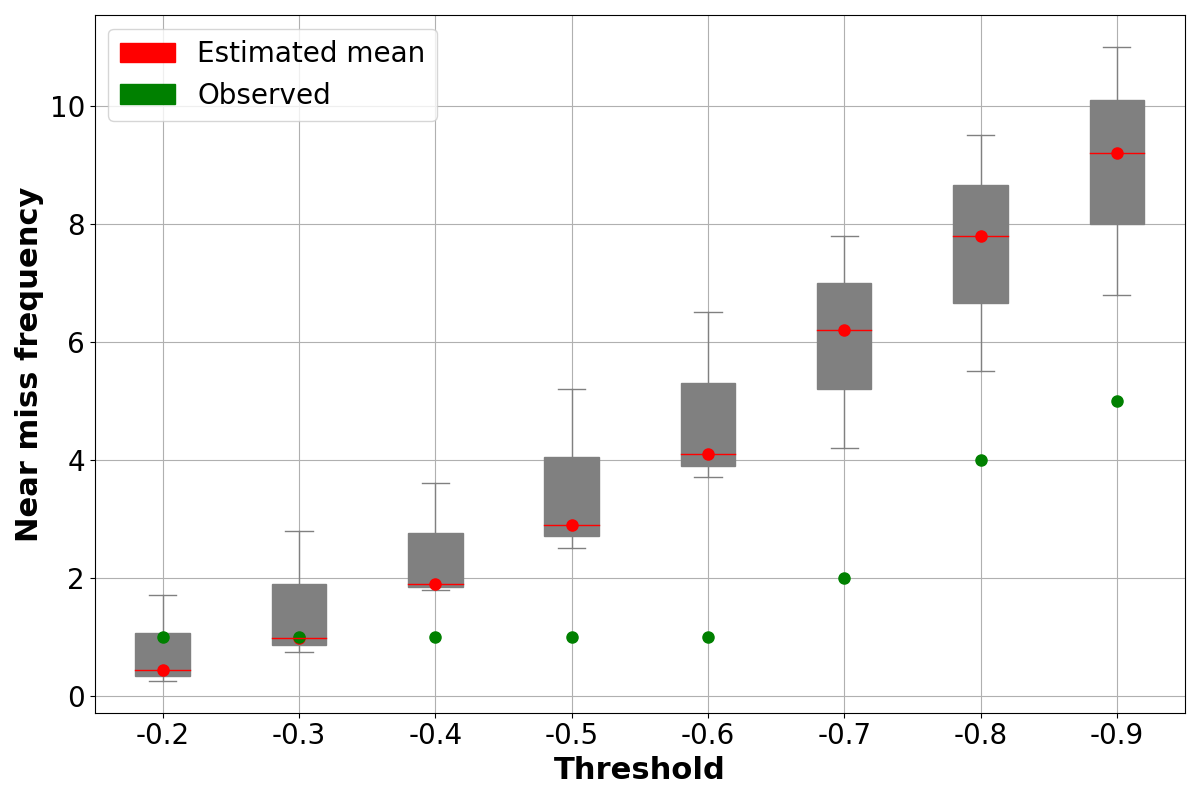}}
    \subcaptionbox{Site-3}{\includegraphics[width=0.4\textwidth]{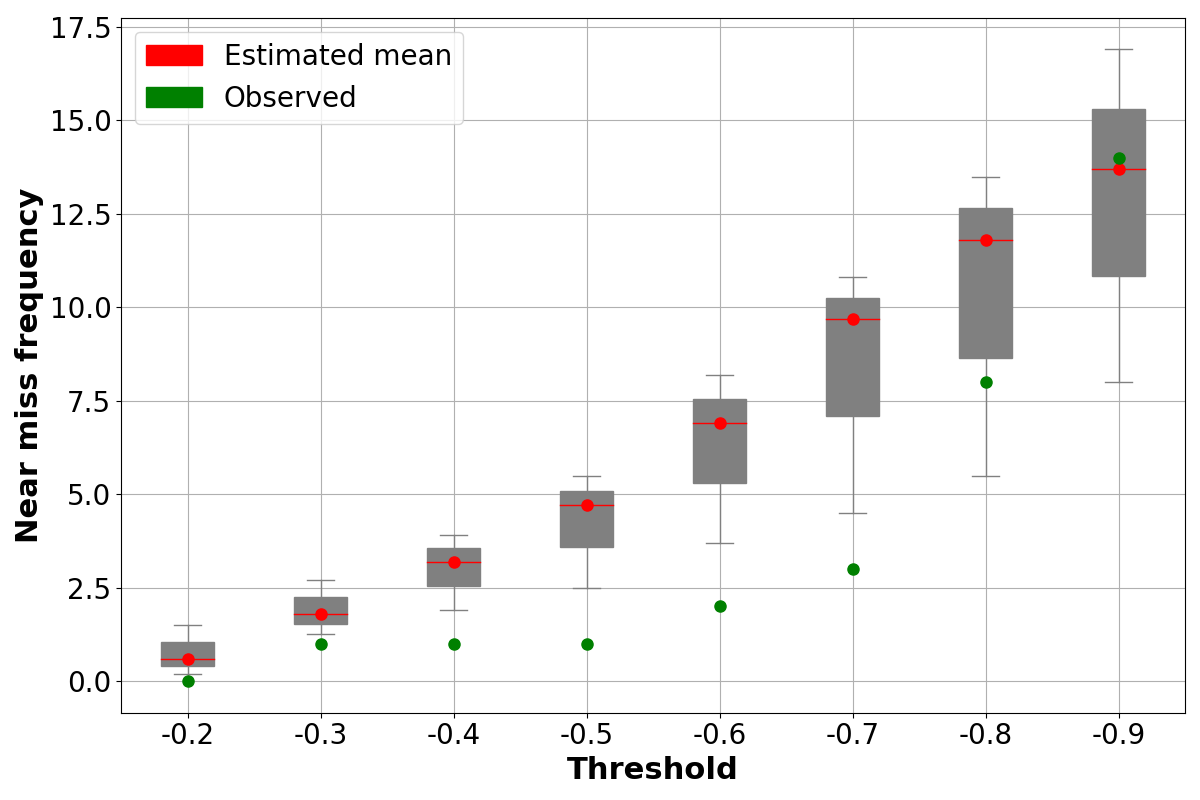}} 
    \subcaptionbox{Site-4}
    {\includegraphics[width=0.4\textwidth]{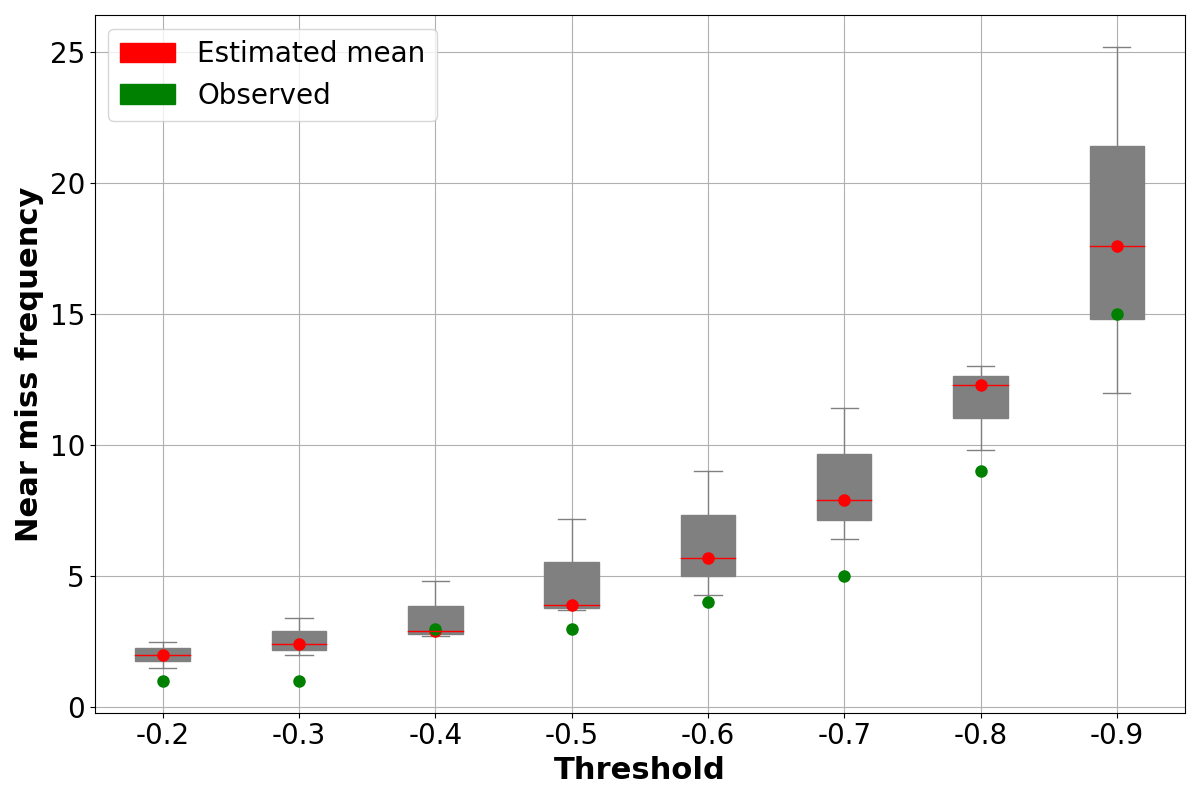}}
    \subcaptionbox{Site-5}{\includegraphics[width=0.4\textwidth]{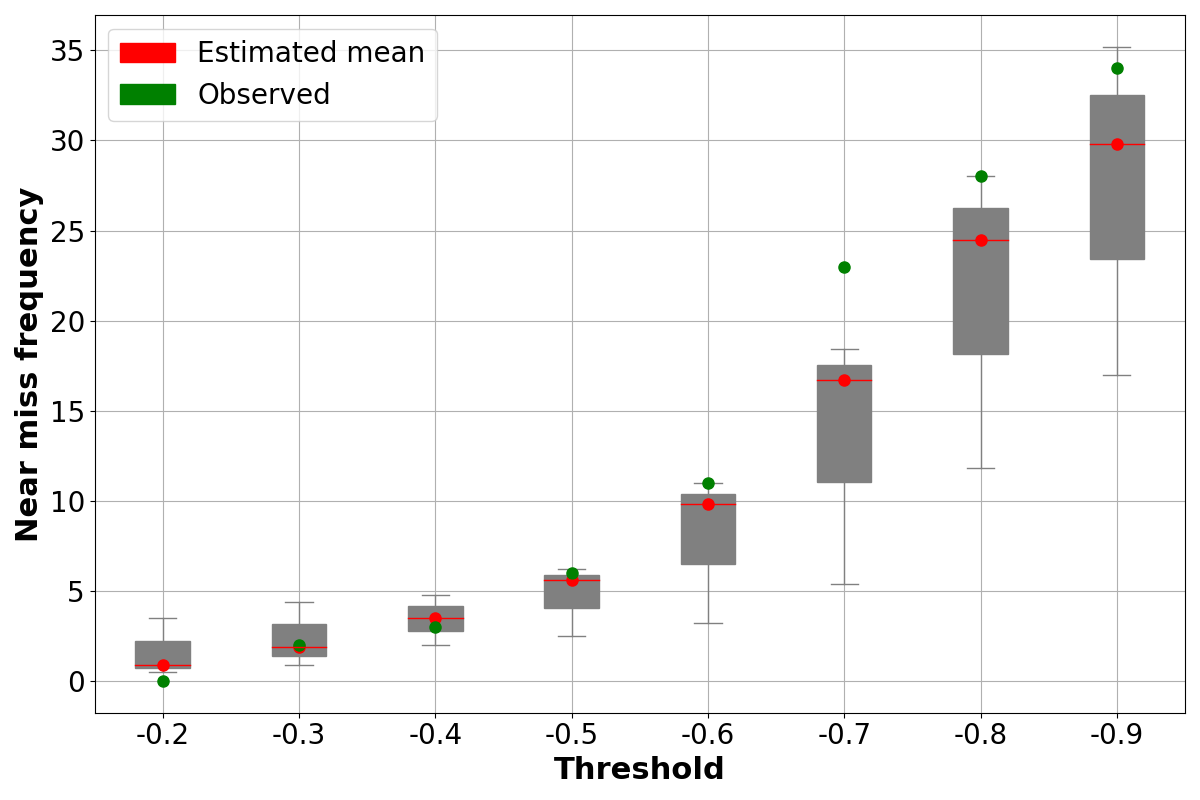}}
    \subcaptionbox{Site-6}{\includegraphics[width=0.4\textwidth]{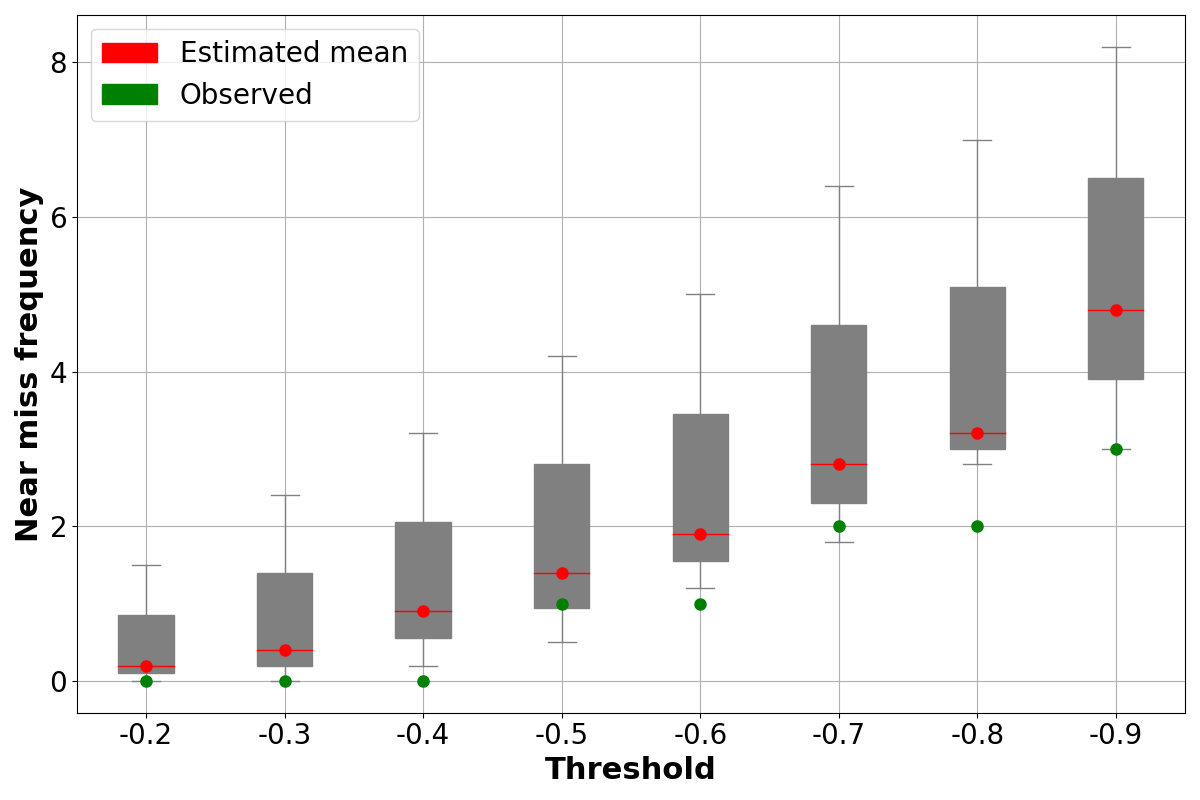}}      
    \caption{Result comparison of observed and estimated near miss using HBSRP model} 
    \label{fig:13}
\end{figure}

Here, \(m\) denotes the site type, \(\lambda\) represents the threshold values (this validation considered: -0.2s, to -0.9s are extreme near misses), and \(k\) is the number of blocks for each site. The parameters \(\mu_n\), \(\vartheta_n\), and \(\xi_n\) are the location, scale, and shape parameters of the HBSRP model distribution, respectively. Table \ref{tab:6} comprehensively compares observed and estimated near-miss incidents across six different sites and thresholds ranging from -0.2s to -0.9s, highlighting key insights into the model's predictive performance. Each cell in the table displays the estimated near misses, followed by the observed near misses in parentheses. The model overestimates near misses at severe thresholds, suggesting that the model may be overly sensitive at these severe thresholds, leading to higher predicted values than found near misses using Eqn. \ref{eq5} and \ref{eq6}. Conversely, the model underestimates the near misses at less severe thresholds, which indicates that the model is restricted to accurately predicting the number of near misses under less severe conditions. Figure \ref{fig:13} shows that the estimated mean values for most sites are generally higher than the observed value. The confidence intervals indicate the range within which the true number of near misses is expected to lie in 95\% confidence. The model seems to provide reasonable estimates, although there are instances where the observed near-misses fall outside the estimated confidence intervals.

\begin{table}[h]
    \centering
    \caption{Estimated and observed near misses at different sites and thresholds}
    \begin{tabular}{ccccccc}
        \toprule
        \textbf{${\lambda}$ \text{(sec)}} & \text{Site} 1       & \text{Site} 2       & \text{Site} 3       & \text{Site}4       & \text{Site} 5       & \text{Site} 6       \\
        \midrule
        -0.2 & 0.67 (\textbf{0})   & 0.43 (\textbf{1})   & 0.6 (\textbf{0})    & 2.0 (\textbf{1})    & 0.9 (\textbf{0})    & 0.2 (\textbf{0})    \\
        -0.3 & 1.3 (\textbf{1})    & 0.98 (\textbf{1})   & 1.8 (\textbf{1})    & 2.4 (\textbf{1})    & 1.9 (\textbf{2})    & 0.4 (\textbf{0})    \\
        -0.4 & 2.5 (\textbf{2})    & 1.9 (\textbf{1})    & 3.2 (\textbf{1})    & 2.5 (\textbf{3})    & 3.5 (\textbf{3})    & 0.9 (\textbf{0})    \\
        -0.5 & 3.9 (\textbf{3})    & 2.9 (\textbf{1})    & 4.7 (\textbf{1})    & 3.9 (\textbf{3})    & 5.6 (\textbf{6})    & 1.4 (\textbf{1})    \\
        -0.6 & 4.7 (\textbf{3})    & 4.1 (\textbf{1})    & 6.9 (\textbf{2})    & 5.7 (\textbf{4})    & 9.8 (\textbf{11})   & 1.9 (\textbf{1})    \\
        -0.7 & 5.7 (\textbf{3})    & 6.2 (\textbf{2})    & 9.7 (\textbf{3})   & 7.9 (\textbf{5})    & 16.7 (\textbf{23})  & 2.8 (\textbf{2})    \\
        -0.8 & 6.6 (\textbf{4})    & 7.8 (\textbf{4})    & 11.8 (\textbf{8})   & 12.3 (\textbf{9})   & 24.5 (\textbf{28})  & 3.2 (\textbf{2})    \\
        -0.9 & 8.2 (\textbf{5})   & 9.2 (\textbf{5})    & 13.7 (\textbf{14})  & 17.6 (\textbf{15})  & 29.8 (\textbf{34})  & 4.8 (\textbf{3})    \\
        \bottomrule
    \end{tabular}
    \label{tab:6}
\end{table}

\bigskip

\section{Discussion}\label{sec 5}

This study proactively analyzes near-miss traffic risks using high-fidelity, 2D indicator data derived from Waymo AV sensors, which capture granular vehicle trajectories. Employing EVT models, this approach aims to enhance the accuracy of traffic crash risk estimation in mixed traffic environments in real-time, for instance, in the next 3s. Previous studies have predominantly developed frameworks tailored to specific locations and crash types without incorporating vehicle dynamics and 2D movement complexities.

The proposed framework employs hierarchical Bayesian UGEV models. Selecting a suitable near-miss indicator is crucial for precise crash risk estimation and aligning with the study’s objectives. This study uses 2D TTC, developed by Li et al.(\citeyear{li2024beyond}), to extract 2D near-miss events while vehicle complex maneuvers, such as merging, lane changes, and passing, typically occur in arterial networks. Additionally, this TTC allows near-miss events during vehicle interactions across various highway geometries while accounting for their dynamics and fidelity. Unlike most existing literature, this study calculated TTC based on vehicle longitudinal and steering or lateral movements to identify vehicle near-miss extreme events. The UGEV models were developed using BM sampling, a method well-suited for real-time frameworks. Unlike video-based datasets, which often record long durations (5min-144 hrs), this study's Waymo AV sensor dataset covers very short durations, with 20s intervals for each segment. In the BM model, each block should be sufficiently long to capture sufficient events to identify true extremes (\citep{zheng2014freeway}). To address the challenge of limited segment length, each conflicting vehicle pair is treated as an individual block, their minimum 2D TTC represent as extreme near-miss events. This approach reliably captures vehicle driving behavior heterogeneity even in short segments, treating each 20s duration as a natural block for this study. It successfully addresses the dearth of extreme events in short-duration traffic segments and satisfies Zheng et al.'s (\citeyear{zheng2014freeway}) recommendation of at least 30 events required for robust BM sampling-based modeling. Results indicate that the framework yields reasonably precise risk estimates, though some observed severe near-misses fall outside the estimated confidence intervals. This discrepancy may stem from the limited sample size due to the short duration of each segment.

While Waymo AV traverses roadways, their LiDAR sensors collect trajectories on nearby road users. Despite the small sample size due to the short segment duration, the high-resolution data are useful for capturing the dynamic nature of mixed traffic. The study successfully modeled near-miss events, even with the current level of AV market penetration in the observed areas. Future increases in AV market penetration are expected to improve the accuracy and effectiveness of the model. Fu and Sayed (\citeyear{fu2023identification}) found that larger observation sample sizes in the hierarchical Bayesian GEV model led to lower model uncertainty. They proposed a method for identifying an adequate sample size for conflict-based crash risk estimation models. The framework in this study still provides reasonably precise risk estimates, despite the smaller sample size and shorter observation period. Future studies may extend observation periods to improve the precision of risk estimates. Additionally, a comprehensive investigation into how varying observation periods affect risk estimates' accuracy and confidence intervals in mixed traffic environments is warranted. However, caution should be exercised with larger block sizes, as they often result in higher variance in model parameters (\citep{ali2022assessing}). Additionally, future studies may explore other sampling methods for extreme events, such as the peak-over-threshold (POT) approach with either fixed or dynamic thresholds and the r largest order statistics method. Comprehensive comparisons between these methods and the BM approach could help determine which method yields more precise crash risk estimates. 

The study also explored whether the identified extremes fit better with stationary or non-stationary UGEV distributions. Among the four UGEV models estimated, the non-stationary model for location and scale with random parameters (HBSRP) performed best. The non-stationary models were addressed by parameterizing the location and scale parameters and incorporate relevant covariates and random intercepts. Specifically, utilized covariates such as conflicting vehicle speed and acceleration to capture the dynamic nature of the data. Studies have highlighted the importance of incorporating vehicular heterogeneity in crash risk assessment (\citep{kumar2024risk}). This is particularly relevant in traffic crashes, where vehicle speeds and accelerations are critical indicators of surrounding road traffic situations (\citep{fu2021random}).  The model with these covariates outperformed competing models, as local and global GOF measures demonstrated. This superior performance indicates that including these specific covariates significantly enhanced the model's ability to predict crash risks accurately, providing a more precise and reliable assessment of vehicle dynamics and near-miss events. First, a local GOF measure was performed by comparing probability density plots (empirical versus modeled). The results showed that the severe extremes identified by HBSRP follow extreme value distributions, indicating their suitability as an alternative to other modeling approaches.

Since the study dataset is anonymous, there is no definitive ground truth of real extremes or crashes. Therefore, a cross-fold validation approach was used to evaluate the performance of the HBSRP model, splitting the dataset into training and test sets to assess the model’s predictive accuracy. This method allows for estimating expected extreme near-miss events at different severity levels and comparison with observed near-miss frequencies.  Global GOF measures, such as mean near-miss, which may be overly sensitive estimates and confidence intervals, were used to assess the model's performance. The model overestimates near misses at severe thresholds, suggesting it may be overly sensitive. Conversely, the model underestimates near misses at less severe thresholds, indicating a limitation in predicting the number of near misses under these conditions. It was also found that the mean near-miss estimates for BM, fitted to UGEV distributions, fell outside the confidence intervals of the observed near misses at some sites, indicating a potential limitation in the model's accuracy.

In summary, the proposed framework represents substantial potential for active traffic management systems, which can be seamlessly integrated into existing traffic control infrastructure, leveraging big data or similar traffic video camera data to estimate potential collisions using vehicle trajectories. This advancement allows proactive, dynamic interventions such as issuing warnings or activating traffic control measures, which can reduce crash risks at hotspot areas like intersections, ramp merges, etc. The modular and scalable framework makes it adaptable for broader applications, including integration into digital twin environments, broad adaptability across different urban and rural environments, road geometries, and traffic conditions for extensive active safety research. Incorporating large-scale datasets, such as NGSIM, Agroverse, and Wejo, enabling it to be applied across various geographic locations with different traffic characteristics, will enhance the generalizability of the proposed framework. However, there is a need to resolve its computational demand to perform it proactively, particularly in dense urban areas, where, in general, processing high-volume data requires significant computational power. Current-time advancements in edge computing and cloud-based solutions have made real-time processing feasible enough. This study involved a limited set of interactions such as HDV-HDV, HDV-AV, and AV-HDV conflicting pairs; future work will incorporate more detailed vehicle types (e.g., trucks, passenger cars) and specified human driver behaviors, enhancing the framework’s capacity for behavior-embedded safety analysis.

\section{Summary and Conclusions}\label{sec 6}
This study proposes a novel framework for real-time estimating near-miss traffic risk using high-resolution vehicle trajectory data from Waymo autonomous vehicle (AV) sensors. The proposed framework incorporates comprehensive vehicle dynamics in 2D, as outlined in Eqn \ref{eq5}, allowing for the prediction of future vehicle trajectories and identification of potential conflicts between vehicles. Near-miss events are defined based on TTC values, with thresholds set between 0.1s$ \le$ TTC $\le $3s. Extreme near-miss events are extracted using the block maxima approach, as described in Eqn \ref{eq6}, which is then integrated into a hierarchical Bayesian structure combined with Extreme Value Theory (EVT). This integration enables a detailed analysis of extreme near-miss risks by leveraging granular vehicle trajectory data.

The parameters, such as $x$, $y$, $v$, $a$, $\theta$, $\delta$, $L$, and $r$, have influence in shaping extreme near-miss events and potential crashes. By focusing on these parameters, this study provides a more detailed understanding of active traffic safety compared to traditional methods that rely on aggregated data.  Additionally, the use of non-stationary and random parameters in the hierarchical Bayesian structure, namely the HBSRP model, accounts for unobserved heterogeneity across different sites, improving the accuracy of crash risk estimations. This framework was calibrated across six urban arterial sites (San Francisco, Phoenix, and Los Angeles), demonstrating its capability to offer real-time risk estimation over short-duration intervals (20s)(\citep{kamel2023real, kamel2024real, fu2022bayesian,fu2022random}). 

The study also acknowledges the heterogeneous nature of traffic, incorporating varying states of interacting vehicles and their influence on near-miss risk. By including covariates such as speed and acceleration of conflicting vehicles, the HBSRP model captures temporal and spatial variations across different sites, thus improving model precision. The framework's ability to accurately represent extreme events, particularly in modeling the tail behavior of the risk distribution, makes it highly effective in identifying high-risk scenarios. This allows for more nuanced insights into vehicle interactions, enabling more precise risk estimations and proactive safety measures.

However, several limitations need to be considered. First, potential biases in data collection, such as limitations in LiDAR sensor range and temporal aggregation, may impact the accuracy of traffic state estimates and vehicle detection, affecting the identification of near-miss events and the precision of crash risk assessments. Second, the Eqn \ref{eq5} used to calculate 2D TTC assumes constant wheel torque and steering angle over short time intervals to identify potential collisions via spatial overlap. Additionally, the model linearizes certain vehicle behaviors, leading to potential errors from nonlinearity; these can be mitigated through numerical methods. Third, this study did not consider other traffic covariates, including shockwave, platoon ratio, queue length, etc, which are important covariates in traffic risk estimation. Incorporating these variables could further enhance the accuracy and depth of the model’s predictions in future research.

In future work, expanding the framework to incorporate multivariate EVT models with additional 2D risk indicators such as post-encroachment time (PET), deceleration rate to avoid collision (DRAC), and minimum time-to-collision (MTTC) would provide a more comprehensive assessment of crash risk, especially concerning crash severity. Incorporating data on vulnerable road users (e.g., pedestrians and cyclists), as well as near-miss event observations from different times, days, and seasons, could significantly enhance the model's generalizability across diverse traffic environments. Future studies could also examine the influence of road geometry and infrastructure characteristics on near-miss events, contributing to the development of more effective countermeasures to improve road safety. Moreover, a comparative analysis between this framework and traditional crash prediction models—such as variants of the Negative Binomial-Lindley (NB-L) model, using other datasets like Agroverse and Wejo (not anonymous), could offer valuable insights into model performance, including prediction accuracy, data requirements, and responsiveness to dynamic traffic conditions. Such comparisons would further validate the effectiveness of this real-time risk estimation framework in proactively managing traffic safety. As the penetration of AVs in mixed-traffic environments increases, access to more extensive datasets will enable more comprehensive safety research.  Ultimately, this proactive approach has the potential to contribute significantly to the Vision Zero initiative, aiming to eliminate all traffic fatalities and serious injuries while promoting safe and equitable mobility for all.

\bigskip
\bigskip

\subparagraph{\textbf{Acknowledgement:}}
This research is funded by Federal Highway Administration (FHWA) Exploratory Advanced Research 693JJ323C000010. The results do not reflect FHWA's opinions.

\bigskip
\subparagraph{\textbf{Disclaimer:}}
The results presented in this document do not necessarily reflect those from the Federal Highway Administration.

\bigskip
\subparagraph{\textbf{Data availability:}}
Data and code will be made available on reasonable request.

\printcredits

\bibliographystyle{cas-model2-names}



\end{document}